\documentclass[11pt]{article}
\usepackage{lmodern,eurosym,ae,tabularx,amssymb,url,epsf,epsfig,amsmath,amsfonts,graphicx,dsfont}
\font\dsrom=dsrom10 scaled 1200
\usepackage[latin1]{inputenc}
\usepackage[english]{babel}
\usepackage{amsmath, amsfonts,amssymb}
\usepackage{amsthm}
\usepackage[T1]{fontenc}
\usepackage{graphics}
\usepackage{graphicx}
\usepackage{geometry}
\newenvironment{pf}{\ \\ {\bf Proof: }}{\hfill\mbox{$\diamond$}\medskip}

\newtheorem{rem}{Remark}[section]
\newtheorem{thm}{Theorem}[section]
\newtheorem{lem}{Lemma}[section]

\newtheorem{cor}{Corollary}[section]

\newtheorem{definition}{Definition}[section]

\newtheorem{prop}{Proposition}[section]

\renewcommand{\epsilon}{\varepsilon}
\newcommand{\be}{\begin{equation*}}
\newcommand{\ben}{\begin{equation}}
\newcommand{\ee}{\end{equation*}}
\newcommand{\een}{\end{equation}}
\renewcommand{\qed}{\hfill\mbox{$\diamond$}\medskip}

\newcommand{\R}{\mathbb{R}}
\renewcommand{\P}{\mathbb{P}}
\newcommand{\E}{\mathbb{E}}
\newcommand{\C}{\mathcal{C}}
\newcommand{\N}{\mathbb{N}}
\newcommand{\D}{\mathcal{D}}

\renewcommand{\phi}{\varphi}
\newcommand{\B}{\mathcal{B}}
\newcommand{\F}{\mathcal{F}}
\newcommand\ind[1]{\textrm{\dsrom{1}}_{#1}}

\begin{document}

\title{Exercise Boundary of the American Put Near Maturity
 in an Exponential Lévy Model}

\author{Damien {\sc Lamberton} \footnote{Universit\'e Paris-Est,
Laboratoire d'analyse et de math\'ematiques appliqu\'ees,
UMR CNRS~8050,
5, Bld Descartes,  F-77454 Marne-la-Vall\'ee Cedex 2, France.  E-mail: {\tt
damien.lamberton@univ-mlv.fr}} \and
Mohammed \sc{Mikou}\footnote{École Internationale des Sciences
du Traitement de l'Information,
Laboratoire de math\'ematiques,
avenue du Parc, 95011 Cergy-Pontoise Cedex.  E-mail: {\tt
mohammed.mikou@eisti.eu} }}

\maketitle
\begin{abstract}
We study  the behavior of the critical price of an American put option
near maturity in the exponential Lévy model when the underlying stock
pays dividends at a continuous rate. In particular, we prove that, in
situations where the limit of the critical price is equal to the stock price,
the rate of  convergence to the limit is linear if and only if the underlying
L\'evy process has finite variation.
In the case of infinite variation, a variety of rates of convergence can be
observed: we prove that, when the negative part of the L\'evy measure
exhibits an $\alpha$-stable  density near
 the origin, with $1<\alpha<2$,
the convergence rate is ruled by
$\theta^{1/\alpha}|\ln \theta|^{1-\frac{1}{\alpha}}$, where $\theta$ is time
until maturity.
\end{abstract}

\section{Introduction}
The behavior of the exercise boundary of the American put near maturity
is well understood in the Black-Scholes model. In particular,
 Barles-Burdeau-Romano-Samsoen \cite{BBRS} (see also~\cite{DL})
showed that, in the absence of dividends, the distance
between the strike price $K$ and the critical price at time $t$,
which we denote by $b^{BS}(t)$ satisfies
\begin{eqnarray}\label{eq-estim-BS}
\lim_{t\to T}\frac{K-b^{BS}(t)}{\sigma K\sqrt{(T-t)|\ln(T-t)|}}=1,
\end{eqnarray}
where $T$ is the maturity, and $\sigma$ is the volatility
(see also \cite{ChenChadam} for higher order expansions).

The aim of this paper is to study the exercise boundary of the
American put near maturity in exponential L\'evy models.
Note that Pham \cite{pham} proved that the estimate
\eqref{eq-estim-BS} holds in a jump diffusion model
satisfying  some conditions. We will first extend Pham's result to slightly
more general situations and, then,
we  will concentrate on L\'evy processes with
no Brownian part. In a recent paper (see \cite{DM}),
we characterized the limit of the critical price at maturity for general
exponential L\'evy models (see also Levendorskii~\cite{Lev}
for earlier related results). In particular, we proved
that, if the interest rate $r$ and the dividend rate $\delta$
satisfy
\begin{equation}
\label{d+>0}
r-\delta\geq \int \left(e^y-1\right)_+\nu(dy),
\end{equation}
where $\nu$ is the L\'evy measure of the underlying L\'evy process,
the limit of the critical price at maturity is equal to the strike price $K$.
In the present paper, we limit our study to situations where the limit
is equal to $K$.

The early exercise premium formula is crucial in our approach.
This theorem shows  that the American put price is the sum of the European put price and a
term which depends on the free boundary, called the {\it early exercise premium}.
This result was already established by Carr-Jarrow-Myneni \cite{CJM}, Jacka \cite{Jacka} and Kim \cite{Kim}
in the Black-Scholes model, and by Pham \cite{pham} in the jump diffusion model.
In this work, we extend this result to an exponential Lévy model when the
related Lévy process is of type $B$ or $C$ (see the definition p.\pageref{def-X-typ}).

The paper is organized as follows. In
Section $2$, we recall some facts about
 the exponential Lévy model
and the basic properties of the American put price in this model.
In Section $3$, we establish the early exercise premium representation.
In Section $4$, we study the critical price near maturity when
the logarithm of the stock includes a diffusion component and a pure jump
process with finite variation. We show in this case that the estimate
\eqref{eq-estim-BS} remains true.
In the fifth section we prove that the convergence rate of the critical price is
linear with respect to $t$ when the logarithm of the stock is a finite
variation Lévy process (see Theorem~\ref{thm-5.2}).
Section 6 deals with the case when
the logarithm of the stock is an infinite variation Lévy process.
We show in this case that the convergence speed of the critical price is
not linear (see Theorem~\ref{thm-infiniteV}).
 Finally, in Section 7, we study processes with a L\'evy density
which behaves asymptotically like an $\alpha$-stable density
 in a negative neighborhood of the origin with $1<\alpha<2$. In this case,
 the rate of convergence involves time to maturity to a power $1/\alpha$,
 together with a logarithmic term, with exponent $1-\frac{1}{\alpha}$
 (see Theorem~\ref{thm-Stable}).
So, there is a logarithmic factor (as in the Black-Scholes case),
in contrast with the finite variation setting where the behavior is purely
 linear.
 \section{The model}
 \subsection{Lévy processes}
A real Lévy process $X=(X_t)_{t\geq 0}$
is a {c\`adl\`ag\footnote{The sample paths of $X$
are right continuous with left limits}} real valued stochastic process,
starting from $0$, with stationary and independent increments.
The random process $X$ can be interpreted
as the independent superposition of a Brownian
motion with drift and an infinite superposition of independent
(compensated) Poisson processes. More precisely the Lévy-Itô
decomposition (see \cite{sato}) gives the following representation
of $X$
 \begin{eqnarray} \label{eq-decom-X}
  X_t &=& \gamma{t} + \sigma B_t  + Y_t, \quad t\geq 0,
 \end{eqnarray}
where $\gamma$ and $\sigma$ are real constants,
$(B_t)_{t\geq 0}$
is a standard  Brownian motion, and the process $Y$ can be written
in terms of the jump measure $J_X$ of $X$
\begin{equation}\label{Y}
Y_t=\int_0^t\int_{\{|x|>{1}\}}{x}J_{X}(ds,dx)+
    \int_0^t \int_{\{0<{|x|}\leq{1}\}}{x}\tilde{J}_{X}(ds,{dx}), \quad t\geq 0.
\end{equation}
Recall that $J_{X}$ is a Poisson measure on
$\mathbb{R}_+\times(\mathbb{R}\setminus\{0\})$, with intensity $\nu$, and
$\tilde{J}_{X}(dt,dx) = {J}(dt,dx) - dt\nu(dx)$ is the compensated
Poisson measure. The measure
$\nu$ is a positive Radon measure on
$\mathbb{R}
\backslash\{0\}$, called the Lévy measure of $X$, and it satisfies
\begin{eqnarray}\label{nu} \int_{\mathbb{R}}
({{1}\wedge{x^2}})\nu{(dx)}<\infty.
\end{eqnarray}
The Lévy-Ito decomposition entails that the
distribution of $X$ is uniquely determined by $(\sigma^2,\gamma,\nu)$,
which is called the characteristic triplet  of the process $X$. The
characteristic function of $X_t$, for $t\geq 0$, is given by the Lévy-Khinchin
representation (see \cite{sato})
\begin{equation}\label{Levy_Kin}
\mathbb{E} [ e^{iz.X_t}]=
\exp [t\phi(z)],\quad  z\in\mathbb{R},
\end{equation}
 with
 \[
 \phi(z) = -\frac{1}{2}\sigma z^2 + i\gamma.z + \int (e^{izx} -
1 - izx{\mathbf{1}}_{|x|\leq1})\nu{(dx)}.
\]
The Lévy process $X$ is a Markov process and its infinitesimal generator is given by
\begin{eqnarray} \label{troi}
L f(x)&=&\frac{{\sigma}^2}{2}\frac{{\partial}^2f}{\partial{x^2}}(x) + \nonumber
\gamma{\frac{\partial{f}}{\partial{x}}(x)}
\\&&  + \int \left(f(x+y) - f(x) -
y\frac{\partial{f}}{\partial{x}}(x){\mathbf{1}}_{|y|\leq1}\right)\nu{(dy)},
\end{eqnarray}
for every $f\in \mathcal{C}_{b}^2(\R)$, where $\mathcal{C}_{b}^2(\R)$
 denotes the set of all bounded $\mathcal{C}^2$
functions with bounded derivatives.
We recall the following classification of the Lévy process (see \cite{sato}).

\begin{definition}\label{def-X-typ} Let $X$ a real Lévy process with characteristic triplet $(\sigma^2,\gamma,\nu)$.
We say that $X$ is of
\begin{itemize}
  \item \textbf{type $A$}, if $\sigma=0$ and $\nu(\R)<\infty$;
  \item \textbf{type $B$}, if $\sigma=0$, $\nu(\R)=\infty$ and $\int_{|x|\leq1}|x|\nu(\R)<\infty$ ( infinite activity and finite variation);
  \item \textbf{type $C$}, If $\sigma>0$ or $\int_{|x|\leq1}|x|\nu(\R)=\infty$ (infinite variation).
\end{itemize}
\end{definition}

We complete this section with two classical results on L\'evy processes.
The first one concerns the behavior at the origin
(see \cite{sato}, Section 47).
\begin{lem}\label{YV0}Let $Y$ the pure jump process defined in \eqref{Y}.
If $\int_{|x|\leq1}|x|\nu(dx)<\infty$, then

\[\lim_{t\to0^+}\frac{Y_t}{t}=-\int_{|x|\leq1}x\nu(dx),
\] almost surely.

\end{lem}
The other classical result is the so-called {\em compensation formula}
(see \cite{Bertoin}, preliminary chapter). We denote by $\Delta X_t=X_t-X_{t^-}$
the jump of the process $X$ at time $t$.
 \begin{prop}\label{prop-formule-comp}
 Let $X$ be a real Lévy prosess and
 $\Phi$ : $(t,\omega,x)\mapsto\Phi_t^x(\omega)$ a measurable
 nonnegative  function
 on $\R^+\times\Omega\times\R$, equipped with the $\sigma$-algebra
 $\mathcal{P}\otimes \B(\R)$,
 where $\mathcal{P}$ is the predictable $\sigma$-algebra
 on $\R^+\times\Omega$, and  $\B(\R)$ is the Borel $\sigma$-algebra on $\R$.
 We have,
 \begin{eqnarray}\label{eq-prev-fonc}
  \E\left(\sum_{0\leq s<\infty}\ind{\{\Delta X_s\neq0\}}\Phi_s^{\Delta X_s}\right)=\E\left[\int_0^\infty ds\int\nu(dy)\Phi_s^y\right].
  \end{eqnarray}
 \end{prop}

 \begin{rem}\label{rm-formule-comp}
 \rm{The equality \eqref{eq-prev-fonc} remains true
 if the non-negativity assumption on $\Phi_t^x$ is replaced by the
  condition
 $$\E\left[\int_0^\infty ds\int\nu(dy)|\Phi_s^y|\right]<\infty.$$
 }
 \end{rem}

\subsection{The exponential L\'evy model}
In the exponential L\'evy model, the price process
$( S_t)_{t\in[0,T]}$
of the risky asset
is given by
\begin{equation}\label{pre}\
    S_t  = { S_0}e^{(r-\delta)t + X_t},
\end{equation}
where the interest rate $r$, the dividend rate $\delta$ are
nonnegative constants and $(X_t)_{t\in[0,T]}$ is a real Lévy process
with characteristic triplet $(\sigma^2, \gamma, \nu)$. We include
$r$ and $\delta$ in $\eqref{pre}$ for ease of notation.

Under the pricing measure
$\mathbb{P}$, the discounted  dividend adjusted stock price
$(e^{-(r-\delta)t}S_t)_{t\in [0,T]}$ is a martingale, which is
equivalent (see, for instance, \cite{Cotak}),
to the two conditions
\begin{equation}\label{eq-cd-martin} \int_{|x|\geq{1}}e^x\nu(dx)<\infty
     \quad
     \text{and}\quad \frac{\sigma^2}2+ \gamma + \int{(e^x - 1-x\mathbf{1}_{|x|\leq{1}})\nu(dx)} = 0.
    \end{equation}
 We suppose that these conditions are satisfied in the sequel.
We deduce from \eqref{eq-cd-martin} that the
infinitesimal generator defined in $\eqref{troi}$ can be written
as
\begin{equation} \label{l}
Lf(x)=\frac{{\sigma}^2}{2}\left(\frac{{\partial}^2f}{\partial{x^2}} -
\frac{\partial{f}}{\partial{x}}\right)(x) + \int\left(f(x+y) - f(x) - (e^y -
1)\frac{\partial{f}}{\partial{x}}(x)\right)\nu{(dy)}.
\end{equation}

The stock price $( S_t)_{t\in[0,T]}$ is also a Markov process and
$S_t = S_0e^{\tilde{X}_t},$ where  $\tilde{X}$ is a Lévy process
with characteristic triplet $(\sigma^2, r - \delta+ \gamma, \nu)$.
We denote by  $\tilde L$ the infinitesimal generator of $\tilde{X}$.
So, from $\eqref{l}$, we have

\begin{eqnarray} \label{eq-generator}
\tilde{L}f(x)&=&\frac{{\sigma}^2}{2}\frac{{\partial}^2f}{\partial{x^2}}(x)
+ (r -\delta-
\frac{{\sigma}^2}{2})\frac{\partial{f}}{\partial{x}}(x) + \tilde{\mathcal{B}}f(x),
\end{eqnarray}
where
\[
 \tilde{\mathcal{B}}f(x) = \int\nu{(dy)}\left(f(x+y) - f(x) - (e^y -
1)\frac{\partial{f}}{\partial{x}}(x)\right).
\]


\subsection{The American put price}
In this model, the value at time $t$ of an American
put with maturity $T$ and strike price $K$
 is given  by
\begin{eqnarray*} P_t = ess\sup_{\tau\in\mathcal{T}_{t,T}}
\mathbb{E}(e^{-r\tau}\psi(S_\tau)\mid\mathcal{F}_t),
\end{eqnarray*}
where $\psi(x)=(K-x)_+$ and $\mathcal{T}_{t,T}$ denotes the set of
stopping times satisfying $t\leq \tau \leq T$. The filtration
$(\F_t)_{t\geq 0}$ is the usual augmentation of the natural filtration of $X$.
It can be proved (see, for instance, \cite{Mikou}) that
\[
P_t=P(t, S_t),
\]
where,
\begin{equation}\label{eq-put-price}
    P(t,x) =\sup_{\tau\in\mathcal{T}_{0,T -
t}} \mathbb{E}(e^{-r\tau}\psi({S}_\tau^x)),
\end{equation}
with $S^x_t=xe^{(r-\delta)t+X_t}$.
The following proposition follows easily from~\eqref{eq-put-price}.
\begin{prop}
    For $t\in [0,T]$, the function
$x\mapsto P(t,x)$ is non-increasing and convex
on $[0,+\infty)$.

For $x\in [0,+\infty)$, the function $t\mapsto P(t,x)$ is continuous
and  nondecreasing on $[0,T]$.
\end{prop}
Note that we also have $P(t,x)\geq P_e(t,x)$, where
$P_e$ denotes the European put price, defined by
$$P_e(t,x)=\E(e^{-r(T-t)}\psi(S_{T-t}^x)), \quad  (t,x)\in[0,T]\times\R^+.$$

Lamberton and Mikou \cite{DM} showed that the American put price
satisfies a variational inequality in the sense of distributions.
It is more convenient to state this variational inequality after
a logarithmic change of variable. Define
\begin{equation}\label{eq-put-logprice}
    \tilde{P}(t,x)=P(t,e^x), \quad (t,x)\in [0,T]\times \R.
\end{equation}
We have
\[
\tilde{P}(t,x)=\sup_{\tau\in\mathcal{T}_{0,T -
t}} \mathbb{E}(e^{-r\tau}\tilde{\psi}(x+\tilde{X}_\tau)),
\]
where $\tilde{\psi}(x)=\psi(e^x)=(K-e^x)_+$.

\begin{thm}\label{thm-iv}
The distribution $({\partial}_t + \tilde{L} - r )\tilde{P}$ is a
nonpositive measure on $(0,T)\times\mathbb{R}$, and,
on the open set $\tilde C$ we have $({\partial}_t + \tilde{L} - r )\tilde{P}=0$,
where $\tilde C$ is called the continuation region defined by
$\tilde C= \{(t,x)\in(0,T)\times \R \;|\;  \tilde P(t,x)>\tilde{\psi}(x)\}$.
\end{thm}

Lamberton and Mikou \cite{DM} also showed the following proposition.
It will be useful in regularization arguments.

\begin{prop}\label{prop-ComConv}
    If $g\in L^\infty(\R)$, we have,
    for every $\theta\in\D(\R)$,
    $$\tilde L(g*\theta) =\tilde L(g)*\theta,$$
    where $\D({\R})$ is the set of all
    $\mathcal{C}^\infty$ functions with compact support in ${\R}$
\end{prop}

\subsection{The free boundary}
Throughout this paper we will assume that at least one of the
following conditions is satisfied:
\begin{equation}\label{eq-nonsub}
    \sigma\neq 0,\quad \nu((-\infty,0))>0\quad\mbox{or}\quad
    \int_{(0,+\infty)}(x\wedge 1)\nu(dx)=+\infty.
\end{equation}
We then have $\P\left(X_t<A\right)>0$, for all $t>0$ and $A\in \R$, so that
$P_e(t,x)>0$ for every $(t,x)\in [0,T)\times\R_+$.
We will also assume that $r>0$.
The {\em critical price }
 or {\em American critical price } at time $t\in[0,T)$ is defined by

$$b(t) = \inf\{x\geq 0 \; | \;  P(t,x)>\psi(x)\}.$$

Note that, since $t\mapsto P(t,x)$ is nonincreasing, the function
$t\mapsto b(t)$ is nondecreasing.
It follows from \eqref{eq-nonsub} that $b(t)\in [0,K)$.
We obviously have $P(t,x)=\psi(x)$ for $x\in[0,b(t))$ and also
for $x=b(t)$, due to the continuity of $P$ and $\psi$.
We also deduce from the convexity of $x\mapsto P(t,x)$ that

\[
\forall t\in[0,T),\quad \forall x>b(t),\quad P(t,x)>\psi(x).
\]

In other words the continuation region $\tilde C$ can be written as
\[
\tilde C=\{(t,x)\in [0,T)\times [0,+\infty)\;|\; x> \tilde b(t)\},
\]
where $\tilde b(t)=\ln(b(t))$.
The graph of $b$ is called the
{\em exercise boundary} or {\em free boundary}.

It was proved in  \cite{DM} that the function
$b$ is continuous on $[0,T)$, and that $b(t)>0$.
We also recall the following result characterizing  the limit of the critical
price near maturity (see \cite{DM} Theorem~4.4).
\begin{thm}\label{thm-criticalpricenearT}
  Denote
  \[
  d_+= r - \delta-\int(e^x-1)_+\nu(dx).
  \]
   If $d_+\geq 0$, we have $\lim_{t\to T}b(t)=K$.

    If $d_+<0 $, we have $\lim_{t\to T}b(t)=\xi$, where $\xi$ is the unique real number in the
interval $(0,K)$ such that
\begin{eqnarray*}\label{trent1}
\varphi_0 (\xi)= rK,
\end{eqnarray*}
where $\varphi_0$ is the function defined by
\begin{eqnarray*}\varphi_0(x) = \delta x+\int{(xe^{y} - K)_+\nu(dy)} ,
\quad  x\in(0,K).
\end{eqnarray*}
\end{thm}

\section{The early exercise premium formula}
The early exercise premium is the difference $P-P_e$
 between the American  and the European put prices.
 It can be expressed with the help of the exercise boundary.
 This expression can be deduced from the following Proposition,
 which characterizes the distribution
 $(\partial_t+\tilde L-r)\tilde P$ as a bounded measurable function,
 with a simple expression involving the exercise boundary.
\begin{prop}\label{prop-gener}
The distribution $(\partial_t+\tilde L-r)\tilde P$ is given by
\begin{eqnarray}\label{eq-gener}
(\partial_t+\tilde L-r)\tilde P(t,x)=h(t,x), \quad dtdx\mbox{-a.e. on} \,\,(0,T)\times\R^+,
\end{eqnarray}
where $h$ is the function defined by
\begin{eqnarray}\label{h}
h(t,x)=\left[\delta e^x -rK +\int_{\{y>0\}}
\left(\tilde P(t,x+y)-(K-e^{x+y})\right)\nu(dy)\right]\mathbf{1}_{\{x<\tilde b(t)\}},
\end{eqnarray}
with $\tilde b(t)=\ln b(t).$
\end{prop}
\begin{proof}
We know from Theorem~\ref{thm-iv} that, on the open set
\[
\tilde{C}=\{(t,x)\in (0,T)\times \R\;|\; x>\tilde{b}(t)\},
\]
we have $(\partial_t+\tilde L-r)\tilde P(t,x)=0$. On the other hand,
on the open set
\[
\tilde E=\{(t,x)\in(0,T)\times\R^+|\quad x<\tilde b(t)\},
\]
we have $\tilde P=\tilde \psi$, so that, using \eqref{eq-generator}
and $\tilde{\psi}(x)=K-e^x$, we have
\begin{eqnarray*}
(\partial_t+\tilde L-r)\tilde P(t,x)&=&\tilde L\tilde P(t,x)-r(K-e^x)\\
     &=&  \delta e^x-rK+\tilde{\B} \tilde{P}(t,x)\\
     &=& \delta e^x-rK+\int\nu(dy)\left(\tilde
P(t,x+y)-\tilde\psi(x)-(e^y-1)\tilde\psi'(x)\right)\\
&=& \delta e^x-rK+\int\nu(dy)\left(
\tilde{P}(t,x+y)+e^{x+y}-K\right).
\end{eqnarray*}
At this point, we clearly have $(\partial_t+\tilde L-r)\tilde P=h$
on the open sets $\tilde C$ and $\tilde E$. Now, if $\sigma>0$
and $\nu(\R)<\infty$, we know (cf. \cite{zhang})  that the partial derivatives
are locally bounded functions, so that the distribution
$(\partial_t+\tilde L-r)\tilde P=h$ is in fact a locally bounded function,
and, since the complement of $\tilde C\cup\tilde E$ is Lebesgue-negligible,
we deduce \eqref{eq-gener}. Now, observe that $h(t,x)\geq -rK$,
so that we have $-rK\leq (\partial_t+\tilde L-r)\tilde P\leq 0$, at least
if $\sigma>0$ and $\nu(\R)<\infty$. On the other hand, in the general case,
we can approximate the L\'evy process $X$ by a sequence of processes
$X^n$ with finite  L\'evy measures $\nu_n$ and positive Brownian variance
parameters
$\sigma_n^2$, in such a way that the America put prices $P^n$ converge simply
to $P$. We then have convergence of
$(\partial_t+\tilde L-r)\tilde P_n$ to $(\partial_t+\tilde L-r)\tilde P$ in the sense
of distributions, so that the double inequality
$-rK\leq (\partial_t+\tilde L-r)\tilde P\leq 0$ is preserved in the limit.
And we can  conclude as in the special case that \eqref{eq-gener}
is true.
%


\end{proof}

The early exercise premium formula is given by the following theorem.
\begin{thm}\label{thm-eep}
 The American put price $P$ related to a Lévy process $X$ of type  $B$ or $C$ has the following
 representation
 \[
 P(t,x)=P_e(t,x)+e(t,x),
 \]
 where $e$ is the {\it early exercise premium} defined by
 \begin{eqnarray*}e(t,x)=\E\left(\int_0^{T-t} k(t+s,S_s^x)e^{-rs}ds\right),
 \end{eqnarray*}
and the function $k$ is given by
\begin{eqnarray}\label{k}
k(t,x)=\left[rK-\delta x -\int_{\{y>0\}}
       \left(P(t,xe^y)-(K-xe^y)\right)\nu(dy)\right]\mathbf{1}_{\{x<b(t)\}},
\end{eqnarray}
for every $(t,x)\in[0,T)\times\R^+$.
\end{thm}

\begin{proof}
We first extend the definition of $\tilde P$ by setting
\[
\tilde P(t,x)=0, \quad\mbox{for}\quad t\notin [0,T], x\in \R.
\]
Next, we regularize $\tilde P$. Let $(\rho_n)_{n\in\N}$
be a sequence of nonnegative  $C^\infty$ functions on $\R^2$such that,
 for every $n\in\N$,
 $supp(\rho_n)\subset (-1/n,1/n)\times (-1/n,1/n)$
 and $\int_{\R^2}\rho_n=1$.
 Define
$$
\tilde{P}_n(t,x)=(\tilde{P}*\rho_n)(t,x)=\int_{\R^2} \tilde P(t-v,x-y)\rho_n(v,y)dvdy, \quad(t,x)\in \R\times\R.
$$
Note that, for each $n$, the function $\tilde{P}_n$ is $\C^\infty$,
with bounded derivatives, and that we have
\begin{equation}\label{eep1}
\forall (t,x)\in (0,T)\times \R,\quad 0\leq \tilde{P}_n(t,x)\leq K\quad
\mbox{and}\quad \lim_{n\to\infty}\tilde{P}_n(t,x)=P(t,x).
\end{equation}
Now, fix $t$ in the open interval $(0,s)$, and let
\[
f_n(s,y)=\tilde{P}_n(t+s,y),\quad (s,y)\in\R\times\R.
\]
Since $f_n$ is smooth with bounded derivatives, we have for any time
$t_1$,  with $0<t_1<T-t$ and any $x\in\R$,
\[
\E\left(e^{-rt_1}f_n(t_1,x+\tilde{X}_{t_1})\right)
    =f_n(0,x)+\E\left[\int_0^{t_1}e^{-rs}
    \left(\partial_s + \tilde L -r\right)f_n(s,x+\tilde{X}_s)ds\right].
\]
Recall that $(\tilde X_t)_{t\in[0,T]}$
  is   defined by  $\tilde X_t=(r-\delta)t+X_t$, and that $\tilde L$
  is the infinitesimal generator of $\tilde X$.

  We have, using Propositions~\ref{prop-ComConv} and \ref{prop-gener}
  \begin{eqnarray*}
 \left(\partial_s + \tilde L -r\right)f_n&=&
          \rho_n* \left(\partial_s + \tilde L -r\right)\tilde P_n(t+\cdot,\cdot)\\
          &=&\rho_n* h(t+\cdot,\cdot).
\end{eqnarray*}
Note that $-rK\leq h\leq 0$, and it follows from \eqref{h}
that $h$ is continuous on the set
$\{(s,y)\;|\; 0<s<T\mbox{ and } y\neq \tilde{b}(s)\}$.

 Now, since $\tilde X$ is a Lévy process of type $A$ or $B$,
 we have for every   $s>0$ (see \cite{sato})
$$
\P\left(t+x+\tilde X_s= \tilde b(s)\right)=0,
$$
so that, by dominated convergence
\[
\lim_{n\to \infty}\E\left[\int_0^{t_1}e^{-rs}
    \left(\partial_s + \tilde L -r\right)f_n(s,x+\tilde{X}_s)ds\right]
    =\E\left[\int_0^{t_1}e^{-rs}
    h(t+s,x+\tilde{X}_s)ds\right]
\]
On the other hand, using \eqref{eep1}, we have
$\lim_{n\to\infty}f_n(0,x)=\tilde{P}(t,x)$ and
\[
\lim_{n\to\infty}\E\left(e^{-rt_1}f_n(t_1,x+\tilde{X}_{t_1})\right)
  =\E\left(e^{-rt_1}\tilde{P}(t+t_1,x+\tilde{X}_{t_1})\right).
\]
Hence
\[
\E\left(e^{-rt_1}\tilde{P}(t+t_1,x+\tilde{X}_{t_1})\right)=
  \tilde{P}(t,x)+
  \E\left[\int_0^{t_1}e^{-rs}
    h(t+s,x+\tilde{X}_s)ds\right].
\]
Now, take the limit as $t_1\to T-t$, and use the continuity of $\tilde P$
on $[0,T]\times \R$ to derive
\[
\E\left(e^{-r(T-t)}\tilde{P}(T,x+\tilde{X}_{T-t})\right)=
  \tilde{P}(t,x)+\E\left[\int_0^{T-t}e^{-rs}
    h(t+s,x+\tilde{X}_s)ds\right].
\]
We have $P(t,x)=\tilde{P}(t,\ln x)$ and
\[
P_e(t,x)=
\E\left( e^{-r(T-t)}\left(K-xe^{\tilde{X}_{T-t}}\right)_+\right)=
\E\left(e^{-r(T-t)}\tilde{P}(T,\ln x+\tilde{X}_{T-t})\right),
\]
so that
\[
P(t,x)=P_e(t,x)-\E\left[\int_0^{T-t}e^{-rs}
    h(t+s,\ln x+\tilde{X}_s)ds\right],
\]
and the early exercise premium formula follows, using
the equality $k(t,x)=-h(t,\ln x)$.
\end{proof}

\begin{rem}\label{rem-eep}
{\rm It follows from Proposition~\ref{prop-gener}  that
$h\geq -rK\mathbf{1}_{x<\tilde b(t)}$, so that, for $t\in(0,T)$ and
$s\in(0,T-t)$, $\liminf_{n\to\infty}\rho_n*h(t+s,x+\tilde{X}_s)\geq
  -rK\mathbf{1}_{\{x+\tilde{X}_s\leq\tilde b(t+s)\}}$.
Using this inequality, we deduce from the proof of Theorem~\ref{thm-eep}
that (even if $X$ is not of type $A$ or $B$),we have
\[
0\leq P(t,x)-P_e(t,x)\leq rK\E\left(\int_0^{T-t}\mathbf{1}_{\{S^x_s\leq b(t+s)\}}ds\right).
\]
}
\end{rem}

The following result will be useful in our study of
the behavior of the critical price when $\sigma>0$,
and in Section~\ref{sec-stable}.

 \begin{cor}\label{cor-eep}
  For every $t\in[0,T)$, the function $x\mapsto P(t,x)-P_e(t,x)$ is nonincreasing on $\R^+$.
 \end{cor}
 \begin{proof}It suffices to show that the early exercise premium $e(t,x)$
 in Theorem~\ref{thm-eep}
 is a nonincreasing function of $x$.
 This will clearly follow if we prove  that $x\mapsto k(t,x)$ is nonincreasing,
 where $k$ is the function defined in \eqref{k}.
 Note that, due to the convexity of $P(t,.)$, the function
 $x\mapsto P(t,x)-(K-x)$ is  nondecreasing, so that
 $$x\mapsto rK-\delta x -\int_{\{y>0\}}\left(P(t,xe^y)-(K-xe^y)\right)\nu(dy),$$
 is nonincreasing.
\end{proof}

\section{The behavior of the critical price when $\sigma>0$}
We suppose throughout this section that $\sigma>0$ and the Lévy measure
satisfies the following condition, which means that the jump-part of the L\'evy process has finite variation,
or, equivalently,
\begin{eqnarray}\label{nu-cd}
\int_{|x|\leq1} |x|\nu(dx)<\infty.
\end{eqnarray}
We will follow the approach of H. Pham \cite{pham}, who treated the case of a finite L\'evy measure.
This section is divided into two parts. In the fist part, we study some links between
European and American put prices in the exponential Lévy model
and  in the Black-Scholes model.
The second part is devoted to the analysis of the critical price near maturity.

\subsection{Links with the Black-Scholes model}
We associate with the continuous part of the L\'evy process
the following Black-Scholes model  $(S^{BS}_t)_{t\in[0,T]}$, given by
 \[
  S^{BS}_t=S_0e^{(r-\delta-\frac{\sigma^2}{2})t+\sigma B_t}, 0\leq t\leq T.
 \]
Denote by $P^{BS}$ and $P_e^{BS}$,
the American put price and the European put price respectively
in this Black-Scholes model. The following Lemma gives an estimate for the difference $P_e-P_e^{BS}$.
\begin{lem}\label{Pee}
there exists a positive constant $C$ such that, for all  $x\in\R^+$, for $t\in [0,T]$,
$$|P_e(t,x)-P_e^{BS}(t,x)|=Cx(T-t).$$
\end{lem}
\begin{proof} Fix  $t\in[0,T]$ and $x\in\R^+$.
Using the decomposition \eqref{eq-decom-X} and the inequality $|a_+-b_+|\leq |a-b|$,
 we have, with the notation $\theta=T-t$,
\begin{eqnarray*}
\left|P_e(t,x)-P_e^{BS}(t,x)\right|&=&
             \left|\E e^{-r\theta}\left[\left(K-xe^{(r-\delta)\theta+\sigma B_\theta+\gamma\theta+Y_\theta}\right)_+
             -\left(K-xe^{(r-\delta-\frac{\sigma^2}{2})\theta+\sigma B_\theta}\right)_+\right]\right|\\
      &\leq&xe^{-\delta\theta}\E\left(e^{\sigma B_\theta-\frac{\sigma^2}{2}\theta}
                   \left|e^{Y_\theta+(\gamma+\frac{\sigma^2}{2})\theta}-1\right|\right)\\
                   &=&xe^{-\delta\theta}
                   \E
                   \left|e^{Y_\theta+(\gamma+\frac{\sigma^2}{2})\theta}-1\right|,
\end{eqnarray*}
where the last equality follows from the independence of $Y$ and $B$.
It follows from \eqref{Y} and \eqref{nu-cd} that $Y_\theta=\hat{Y}_\theta-\int_{\{|y|\leq 1\}}y\nu(dy)$,
where
\[
\hat{Y}_\theta=\int_0^\theta\int yJ_X(ds,dy)=\sum_{0<s\leq \theta}\Delta X_s,
\]
so that, with the notation $\lambda= \gamma+\frac{\sigma^2}{2}-\int_{\{|y|\leq 1\}}y\nu(dy)$,
\begin{eqnarray*}
\E
                   \left|e^{Y_\theta+(\gamma+\frac{\sigma^2}{2})\theta}-1\right|&=&
                   \E
                   \left|e^{\hat Y_\theta+\lambda\theta}-1\right|\\
                   &\leq&\E
                   \left|e^{\hat Y_\theta+\lambda\theta}-e^{\lambda\theta}\right|+\left|e^{\lambda\theta}-1\right|\\
                   &=&e^{\lambda\theta}\E
                   \left|e^{\hat Y_\theta}-1\right|+O(\theta),
\end{eqnarray*}
so that the Lemma will follow if we prove that $\E\left|e^{\hat Y_\theta}-1\right|=O(\theta)$.
We have
\begin{eqnarray*}
e^{\hat Y_\theta}-1&=&
   \sum_{0<s\leq \theta}e^{\hat Y_s}-e^{\hat Y_{s^-}}\\
     &=&\sum_{0<s\leq \theta}e^{\hat Y_{s^-}}\left(e^{\Delta\hat Y_s}-1\right).
\end{eqnarray*}
Hence, using the compensation formula (cf. Proposition~\ref{prop-formule-comp}),
\begin{eqnarray*}
\E\left|e^{\hat Y_\theta}-1\right|&\leq&
   \E\left(\sum_{0<s\leq \theta}e^{\hat Y_{s^-}}\left|e^{\Delta\hat Y_s}-1\right|\right)\\
   &=&\E\int_0^\theta ds e^{\hat Y_{s^-}}\int \nu(dy)\left|e^{y}-1\right|.
\end{eqnarray*}
Note that $\E\left(e^{\hat Y_s}\right)=e^{-\lambda s}$, so that $\E\int_0^\theta ds e^{\hat Y_{s^-}}=O(\theta)$
and that $\int \nu(dy)\left|e^{y}-1\right|<\infty$ due to \eqref{nu-cd} and \eqref{eq-cd-martin}.
 \end{proof}

The following lemma is already established by  Pham \cite{pham}, Bellamy-Jeanblanc
\cite{BellamyJeanblanc} and Jakubenas \cite{Jakubenas}.
\begin{lem}
 \label{lem-Pgeq}We have $P\geq P^{BS}$.
\end{lem}


Recall that $x\mapsto P(t,x)$ is convexe, and denote by $\partial_x^+P(t,x)$
its right-hand derivative. Note that the function $x\mapsto\partial_x^+P(t,x)$
is right continuous on $\R^+$ for every $t\in[0,T]$. We have the following
result about the asymptotic behavior of $\partial_x^+P$ near maturity.
Recall the notation
$d_+= r - \delta-\int(e^x-1)_+\nu(dx)$.
\begin{lem}\label{limbBS}
If $d_+\geq0$, we have
$$\lim_{t\to T}\partial_x^+P(t,b^{BS}(t))=-1,$$
where $b^{BS}$ denotes the critical price in the Black-Scholes model.
\end{lem}

\begin{proof}Fix $t\in[0,T]$ and $x\in\R^+$.
We know from  Corollary~\ref{cor-eep} that $x\mapsto P(t,x)-P_e(t,x)$ is nonincreasing, so that
\begin{eqnarray*}\label{lim0}
-1\leq \partial_x^+P(t,x)\leq\partial_x^+P_e(t,x),
\end{eqnarray*}
where the first inequality follows from the convexity of $x\mapsto P(t,x)$.

We have, with the notation $\theta =T-t$,
\[
P_e(t,x)=e^{-r\theta}\E\left(K-xe^{(r-\delta)\theta+X_\theta}\right)_+,
\]
so that $\partial^+_xP_e(t,x)=-\E\left(e^{-\delta\theta+X_\theta}
          \mathbf{1}_{\{(r-\delta)\theta+X_\theta<\ln({K}/{x})\}}\right)$ and
\begin{eqnarray*}
\partial^+_xP_e(t,b^{BS}(t))=-\E\left(e^{-\delta\theta+X_\theta}
          \mathbf{1}_{\{(r-\delta)\theta+X_\theta<l(\theta)\}}\right),
\end{eqnarray*}
where $l(\theta)=\ln({K}/{b^{BS}(t)})$. Note that
$(e^{-\delta\theta+X_\theta})_{0\leq \theta\leq 1}$
is uniformly integrable (because $(e^{X_t})_{t\geq 0}$ is a martingale
 and that $\lim_{\theta\downarrow 0}e^{-\delta\theta+X_\theta}=1$
  almost surely. Therefore, in order to prove that
  $\lim_{t\to T}\partial^+_xP_e(t,b^{BS}(t))=-1$, it suffices to show that
$\lim_{\theta\downarrow 0}\P\left((r-\delta)\theta+X_\theta<l(\theta)\right)=1$.
We have
\[
\P\left((r-\delta)\theta+X_\theta<l(\theta)\right)=
     \P\left((r-\delta)\sqrt{\theta}+\frac{X_\theta}{\sqrt{\theta}}<
          \frac{l(\theta)}{\sqrt{\theta}}\right)=1,
\]
and, as $\theta$ goes to $0$, $X_\theta/\sqrt{\theta}$ converges
in distribution to a centered Gaussian with variance $\sigma^2$, so that
the result follows from the fact that
$\lim_{\theta\downarrow 0}\frac{l(\theta)}{\sqrt{\theta}}=\infty$
(as a consequence of \eqref{eq-estim-BS}).

\end{proof}
\subsection{Critical price near maturity for $\sigma^2>0$}
We can now state the main result of this Section.
\begin{thm}\label{thm-vpc-sig}
If $d^+>0$, there exists a positive constant $C$ such that
$$0\leq b^{BS}(t)-b(t)\leq C\sqrt{T-t},$$
when $t$ is close to $T$.
\end{thm}
In view of  \eqref{eq-estim-BS}, we deduce the following Corollary.
\begin{cor}\label{cor-vpc-sig} If $d^+>0$, we have

$$\lim_{t\to T}\frac{b(t)-K}{\sigma K\sqrt{(T-t)|\ln(T-t)|}}=1.$$
\end{cor}
\begin{rem}\rm
 Theorem~\ref{thm-vpc-sig} was proved by Pham \cite{pham} in the case of
 a finite L\'evy measure. Our proof is merely an extension of that of \cite{pham}.
 The only difference is in the argument for proving that $P-P_e$ is a non-increasing
 function of the stock price, which in \cite{pham} was based on the maximum principle. 
 \end{rem}
\noindent{\bf Proof of Theorem~\ref{thm-vpc-sig}: }
We first deduce from Lemma~\ref{lem-Pgeq} that $b^{BS}\geq b$.
In order to derive an upper bound for $b^{BS}-b$, we proceed as follows.

We first deduce from Remark~\ref{rem-eep} that, for $0\leq t< T$, $x\geq 0$,
\[
P(t,x)-P_e(t,x)\leq rK(T-t),
\]
so that, using  the inequality $P^{BS}\geq P_e^{BS}$ and Lemma~\ref{Pee},
\begin{eqnarray*}
P(t,x)-P^{BS}(t,x)&\leq &P(t,x)-P_e(t,x)+P_e(t,x)-P_e^{BS}(t,x)\\
         &\leq &(rK+Cx)(T-t).
\end{eqnarray*}
Since $b^{BS}(t)\leq K$, we deduce that, for $t\in [0,T]$ and $x\in [b(t),b^{BS}(t)]$,
\[
P(t,x)-P^{BS}(t,x)\leq \hat{C}(T-t),
\]
with $\hat{C}=(r+C)K$.

On the other hand, we have
\begin{eqnarray*}
P(t,b^{BS}(t))-P^{BS}(t,b^{BS}(t))&=&P(t,b^{BS}(t))-(K-b^{BS}(t))\\
       &=&P(t,b^{BS}(t))-P(t,b(t)) + (b^{BS}(t))-b(t))\\
       &=&P(t,b^{BS}(t))-P(t,b(t)) -(b^{BS}(t))-b(t))\frac{\partial P}{\partial x}(t,b(t)),
\end{eqnarray*}
where the last equality follows from the {\em smooth fit} property (which is valid
because the L\'evy process $X$ has infinite variation, see~\cite{LambertonMikou,Mikou}).

Since the function $x\mapsto P(t,x)$ is convex, its second derivative is a measure on $[0,+\infty)$
and we have
\[
P(t,b^{BS}(t))-P(t,b(t)) -(b^{BS}(t))-b(t))\frac{\partial P}{\partial x}(t,b(t))=
   \int_{b(t)}^{b^{BS}(t)}\frac{\partial^2 P}{\partial x^2}(t,d\xi)(\xi -b(t)).
\]
Hence
\begin{equation}\label{sigma0}
\int_{b(t)}^{b^{BS}(t)}\frac{\partial^2 P}{\partial x^2}(t,d\xi)(\xi -b(t))\leq \hat{C}(T-t).
\end{equation}
We will now deduce a lower bound for $\partial^2 P/\partial x^2$ from the variational inequality.
Indeed, in the open set $\tilde{C}=\{(t,x)\in (0,T)\times \R\;|\; x>\tilde{b}(t)\}$, we have
\[
\left(\tilde{L}-r\right)\tilde P=-\partial_t \tilde{P}.
\]
Since $\tilde{P}$ is a non-increasing function of time, we deduce that $\tilde{L}\tilde{P}\geq 0$ on
$\tilde{C}$. Therefore, going back to the function $P$, we have, for any $t\in (0,T)$, and for $x>b(t)$
\begin{equation}\label{sigma1}
\frac{\sigma^2x^2}{2}\frac{\partial^2 P}{\partial x^2}(t,x)+(r-\delta)x\frac{\partial P}{\partial x}(t,x)
    +\int \left(P(t,xe^y)-P(t,x)-x(e^y-1)\frac{\partial P}{\partial x}(t,x)\right)\nu(dy)\geq 0.
\end{equation}
Note that this inequality holds in the sense of distributions on the open interval $(b(t),+\infty)$
and that $\partial^2 P/\partial x^2$ is to be interpreted as a measure.
Since $x\mapsto P(t,x)$ is non-increasing, we have for $y>0$, $P(t,xe^y)\leq P(t,x)$,
so that
\begin{eqnarray}\label{sigma2}
\int_{y>0} \left(P(t,xe^y)-P(t,x)-x(e^y-1)\frac{\partial P}{\partial x}(t,x)\right)\nu(dy)
   &\leq &-x \frac{\partial P}{\partial x}(t,x)\int_{y>0}(e^y-1)\nu(dy).
\end{eqnarray}
On the other hand, for $y<0$, we have, due to the Lipschitz property of $P(t,\cdot)$
\begin{eqnarray*}
P(t,xe^y)-P(t,x)-x(e^y-1)\frac{\partial P}{\partial x}(t,x)&\leq&2x\left(1-e^y\right).
\end{eqnarray*}
Moreover, if $xe^y<b(t)$,
\begin{eqnarray*}
P(t,xe^y)-P(t,x)-x(e^y-1)\frac{\partial P}{\partial x}(t,x)&=&
      K-xe^y-P(t,x)-x(e^y-1)\frac{\partial P}{\partial x}(t,x)\\
      &\leq &K-xe^y-(K-x)-x(e^y-1)\frac{\partial P}{\partial x}(t,x)\\
      &=&x(1-e^y)\left(1+\frac{\partial P}{\partial x}(t,x)\right).
\end{eqnarray*}
Hence
\begin{eqnarray}
\!\!\!\!\!\int_{y<0} \left(P(t,xe^y)-P(t,x)-x(e^y-1)\frac{\partial P}{\partial x}(t,x)\right)\nu(dy)
  &\!\!\!\!\!\leq \!\!\!\!\!&
     2x\int_{[\ln(b(t)/x), 0)} \left(1-e^y\right)\nu(dy)\nonumber\\
        &&   \!\!\!\!\!\!\!\!\!\!\!+x\!\!\int_{\{y<0\}}\!\!\!(1-e^y)\nu(dy)\left(1+\frac{\partial P}{\partial x}(t,x)\right).\label{sigma3}
\end{eqnarray}
Putting   \eqref{sigma2} and \eqref{sigma3} together, we get, for $x$ in the open interval $(b(t),b^{BS}(t))$,
\begin{eqnarray*}
\int \left(P(t,xe^y)-P(t,x)-x(e^y-1)\frac{\partial P}{\partial x}(t,x)\right)\nu(dy)
&\leq &-x \frac{\partial P}{\partial x}(t,x)\int(e^y-1)_+\nu(dy)+\epsilon(t,x),
\end{eqnarray*}
with
\[
\epsilon(t,x)=2x\int_{[\ln(b(t)/x), 0)} \left(1-e^y\right)\nu(dy)+
       x\int_{\{y<0\}}(1-e^y)\nu(dy)\left(1+\frac{\partial P}{\partial x}(t,x)\right).
\]
Note that if $x\in(b(t),b^{BS}(t))$, we have, by convexity,
$\frac{\partial P}{\partial x}(t,x)\leq \frac{\partial P}{\partial x}(t,b^{BS}(t))$, so that $\epsilon(t,x)\leq \eta(t)$, where
\[
\eta(t)=2b^{BS}(t)\int_{[\ln(b(t)/b^{BS}(t)), 0)} \left(1-e^y\right)\nu(dy)+b^{BS}(t)
     \int_{\{y<0\}}(1-e^y)\nu(dy)\left(1+\frac{\partial P}{\partial x}(t,b^{BS}(t))\right)
\]
Going back to \eqref{sigma1}, we deduce
\begin{eqnarray*}
\frac{\sigma^2x^2}{2}\frac{\partial^2 P}{\partial x^2}(t,x)&\geq& -\left(r-\delta -\int (e^y-1)_+\nu(dy)\right)
        x\frac{\partial P}{\partial x}(t,x)
                                                                         -\eta(t)\\
                                                                         &=&
                                                                         -d_+x\frac{\partial P}{\partial x}(t,x)-\eta(t),
\end{eqnarray*}
so that, using the convexity again,
\begin{eqnarray*}
\frac{\partial^2 P}{\partial x^2}(t,x)&\geq &-\frac{2d_+}{x\sigma^2}\frac{\partial P}{\partial x}(t,x)-
             2\frac{\eta(t)}{\sigma^2 x^2}\\
             &\geq &-\frac{2d_+}{K\sigma^2}\frac{\partial P}{\partial x}(t,b^{BS}(t))-
             2\frac{\eta(t)}{\sigma^2 K^2},\quad b(t)<x<b^{BS}(t).
\end{eqnarray*}
We deduce from this inequality together with \eqref{sigma0}
\[
\hat{C}(T-t)\geq \alpha(t)\left(b^{BS}(t)-b(t)\right)^2,
\]
where
\[
\alpha(t)=-\frac{d_+}{K\sigma^2}\frac{\partial P}{\partial x}(t,b^{BS}(t))-
             \frac{\eta(t)}{\sigma^2 K^2}.
\]
It follows from  Lemma~\ref{limbBS} and \eqref{nu-cd} that $\lim_{t\to T}\alpha(t)=\frac{d_+}{K\sigma^2}$,
so that, under the condition $d_+>0$,
\[
\limsup_{t\to T}\frac{b^{BS}(t)-b(t)}{\sqrt{T-t}}<\infty.
\]
\qed
\section{The critical price near maturity in a finite variation Lévy model}
\label{SectionFV}
Throughout this section, we suppose that  $X$ is a  Lévy process with finite variation,
or, equivalently,
$$
\sigma=0\quad\text{and}\quad\int_{|x|\leq1}|x|\nu(dx)<\infty.
$$
The decomposition  \eqref{eq-decom-X} can then be written as follows
\begin{eqnarray}\label{eq-decom-X-vf}
X_t=\gamma_0 t + \sum_{0<s\leq t}\Delta X_s,\quad t\geq 0,
\end{eqnarray}
where $\gamma_0=:\gamma-\int_{|x|\leq1}x\nu(dx)$. Note that, due to the
martingale condition~\eqref{eq-cd-martin}, we have
\begin{eqnarray}\label{eq-cd-martin-vf}
\gamma_0=-\int (e^y-1) \nu(dy).
\end{eqnarray}
This section is divided into two parts.
In the first part, we introduce what we call the {\em European} critical price,
namely, the stock price value for which the American put price is equal
to its intrinsic value, and we characterize its  behavior near maturity.
 In the second part, we analyze the  difference between the {\em European}
 and {\em American} critical prices and deduce the behavior
of the American critical price.

\subsection{The European critical price}
For each $t$ in the interval $[0,T)$, we define the
{\em European critical price} at time $t$ by
$$
b_e(t)=\inf\{x\in\R^+; \quad P_e(t,x)>\phi(x)\}\quad\forall t\in[0,T).
$$
Note that, since $P_e(t,K)>0$ and $P_e(t,0)=K e^{-r(T-t)}$, we have
$0<b_e(t)<K$. Using the convexity of $P_e(t,\cdot)$, one can see that
$b_e(t)$ is the only real number in the interval $(0,K)$ satisfying the
equality $P_e(t,b_e(t))=K-b_e(t)$.
Recall from Lemma~\ref{lem-Pgeq} that $P\geq P_e$,
so that we have $b\leq b_e(\leq K)$, and
it follows from Theorem~\ref{thm-criticalpricenearT} that,
if $d^+\geq0$,  we have $\lim_{t\to T} b_e(t)=\lim_{t\to T} b(t)=K$.
The following result characterizes the rate of convergence of $b_e(t)$
to $K$.

\begin{thm}\label{thm-vpce-vf}
If $d^+>0$, we have
$$
\lim_{t\to T}\frac{1}{T-t}\left(\frac{K}{b_e(t)}-1\right)=\int(e^y-1)_-\nu(dy).
$$
\end{thm}
\begin{proof}
Starting from the equality $P_e(t,b_e(t))=K-b_e(t)$, we have,
with the notation $\theta=T-t$,
\begin{eqnarray*}
K-b_e(t)&=&\E(e^{-r\theta}(K-b_e(t)e^{(r-\delta)\theta+X_\theta})_+)\\
             &=&e^{-r\theta}K-b_e(t)\E e^{-\delta\theta+X_\theta}
                +\E(e^{-r\theta}(K-b_e(t)e^{(r-\delta)\theta+X_\theta})_-)\\
        &=&e^{-r\theta}K-b_e(t)e^{-\delta\theta}+\E(e^{-r\theta}(b_e(t)e^{(r-\delta)\theta+X_\theta}-K)_+),
\end{eqnarray*}
Dividing both sides by $b_e(t)$, we get
\begin{eqnarray*}
\frac K {b_e(t)}(1-e^{-r\theta})+e^{-\delta\theta}-1=
  \E\left[e^{-r\theta}\left(e^{(r-\delta)\theta+X_\theta}-\frac{K}{b_e(t)}\right)_+
     \right].
 \end{eqnarray*}
Note that, since $\lim_{t\to T}b_e(t)=K$,
\[
\frac K {b_e(t)}(1-e^{-r\theta})+e^{-\delta\theta}-1=(r-\delta)\theta +o(\theta).
\]
Therefore, using the decomposition \eqref{eq-decom-X-vf},
\begin{eqnarray}
(r-\delta)\theta&=&
\E\left(e^{(r-\delta)\theta+X_\theta}-
           \frac{K}{b_e(t)}\right)_+ +o(\theta)\nonumber\\
&=&\E\left(e^{(r-\delta+\gamma_0)\theta+Z_\theta}-\frac{K}{b_e(t)}\right)_+
                   +o(\theta),\label{5.1*}
\end{eqnarray}
with the notation
\[
Z_t=\sum_{0<s\leq t}\Delta X_s, \quad t\geq 0.
\]
We have
\begin{eqnarray}
\E\left(e^{(r-\delta+\gamma_0)\theta+Z_\theta}-\frac{K}{b_e(t)}\right)_+
        &=& \E\left(e^{Z_\theta}[1+(r-\delta+\gamma_0)\theta]-
           \frac{K}{b_e(t)}\right)_++o(\theta) \nonumber\\
        &=& \E\left(e^{Z_\theta}+(r-\delta+\gamma_0)\theta-
           \frac{K}{b_e(t)}\right)_++o(\theta),\nonumber
\end{eqnarray}
where the last equality follows from the fact that
$\lim_{\theta\to0}\E|e^{Z_\theta}-1|=0$. Going back to \eqref{5.1*},
we deduce
\begin{eqnarray}\label{eq-cpe-4}
(r-\delta)\theta =\E(f_\theta(Z_\theta)) + o(\theta),
 \end{eqnarray}
where the function $f_\theta$ is defined by
\[
f_\theta(x)=(e^{x}-1-\tilde\zeta(\theta))_+,\quad
    x\in\R,
\]
with
\[
\tilde\zeta(\theta)=\frac{K}{b_e(t)}-1-(r-\delta+\gamma_0)\theta.
\]
Since the process $Z$ is the sum of its jumps, we have,
using the compensation formula (see Proposition~\ref{prop-formule-comp}),
\begin{eqnarray*}
\E(f_\theta(Z_\theta)) &=& f_\theta(0)+
\E\left(\sum_{0<s\leq\theta}[f_\theta(Z_s)-f_\theta(Z_{s^-})]\right)\\
        &=& f_\theta(0)+
         \E\left(\int_0^\theta ds\int(f_\theta(Z_s+y)-f_\theta(Z_s))\nu(dy)\right)\\
        &=&\left((r-\delta+\gamma_0)\theta-\zeta(\theta)\right)_+
             +\int_0^\theta ds\int\nu(dy)\E\left(f_\theta(Z_s+y)
                  -f_\theta(Z_s)\right),
\end{eqnarray*}
with
\[
\zeta(\theta)=\tilde\zeta(\theta)+(r-\delta+\gamma_0)\theta=
       \frac{K}{b_e(t)}-1.
\]
Note that, since $\lim_{\theta\downarrow 0}\tilde{\zeta}(\theta)=0$,
we have, for any fixed $y\in\R$,
\begin{eqnarray*}
\lim_{\theta\downarrow 0}\frac{1}{\theta}\int_0^\theta ds
\E\left(f_\theta(Z_s+y)\right)&=&
      \lim_{\theta\downarrow 0}\frac{1}{\theta}\int_0^\theta ds
\E\left(e^{Z_s+y}-1-\tilde{\zeta}(\theta)\right)_+\\
&=& \lim_{\theta\downarrow 0}\frac{1}{\theta}\int_0^\theta ds
\E\left(e^{Z_s+y}-1\right)_+\\
&=&\left(e^y-1\right)_+,
\end{eqnarray*}
where the last equality follows from the fact that $\lim_{s\to 0}e^{Z_s}=1$
in $L_1$.

On the other hand, we have
\begin{eqnarray*}
\frac{1}{\theta}\int_0^\theta ds
\E\left(f_\theta(Z_s+y)-f_\theta(Z_s)\right)&\leq &
           \frac{1}{\theta}\int_0^\theta ds\E\left(e^{Z_s}\left|e^y-1\right|\right)\\
           &=&\frac{1}{\theta}\int_0^\theta ds
          e^{-\gamma_0s}\left|e^y-1\right|\\
           &\leq&\frac{e^{|\gamma_0|\theta}-1}{|\gamma_0|\theta}
          \left|e^y-1\right|.
\end{eqnarray*}
Since
$\displaystyle\sup_{0<\theta<1}
    \frac{e^{|\gamma_0|\theta}-1}{|\gamma_0|\theta}<\infty$
    and $\int  \left|e^y-1\right|<\infty$, we deduce, by dominated convergence
    that
    \[
    \lim_{\theta\downarrow 0}\frac{1}{\theta}
    \int_0^\theta ds
\E\left(f_\theta(Z_s+y)-f_\theta(Z_s)\right)=\int  \left(e^y-1\right)_+.
    \]
We can now rewrite
\eqref{eq-cpe-4} as
\[
(r-\delta)\theta=((r-\delta+\gamma_0)\theta-\zeta(\theta))_++\theta
      \int(e^{y}-1)_+\nu(dy)+o(\theta),
\]
so that
\[
d^+\theta =\left(r-\delta-\int (e^{y}-1)_+\nu(dy)\right)\theta=  ((r-\delta+\gamma_0)\theta-\zeta(\theta))_+  +o(\theta).
\]
Since $d^+>0$,
we must have $(r-\delta+\gamma)\theta-\zeta(\theta)>0$
for $\theta$ close to  $0$. Hence
\begin{eqnarray*}
\lim_{\theta\downarrow 0}\frac{\zeta(\theta)}\theta&=&\gamma_0+
        \int (e^{y}-1)_+\nu(dy)\\
        &=&
\int(e^y-1)_-\nu(dy),
\end{eqnarray*}
where the last equality follows from \eqref{eq-cd-martin-vf}.
\end{proof}

\subsection{The behavior of the critical price}
We are now in a position to prove the main result of  this section.

\begin{thm}\label{thm-5.2}
If $d^+>0$, we have
$$
\lim_{t\to T}\frac{1}{T-t} \left(\frac{K}{b(t)}-1\right)=\int(e^y-1)_-\nu(dy).
$$
\end{thm}

\begin{proof}
In view of Theorem~\ref{thm-vpce-vf}, it suffices to prove that
\[
\lim_{t\to T}\frac{b_e(t)-b(t)}{(T-t)}=0.
\]
Recall that $b_e\geq b$ and, from Remark~\ref{rem-eep},
we have
\begin{eqnarray}\label{eq-vitess-b0}
0\leq P(t,x)-P_e(t,x)\leq rK\E\left(\int_0^{T-t}\mathbf{1}_{\{S_s^x\leq b(t+s)\}}ds\right)
,\quad (t,x)\in[0,T)\times\R^+.
\end{eqnarray}
From the equality $P_e(t,b_e(t))=K-b_e(t)$
and the convexity of $P(t,\cdot)$, we deduce
\begin{eqnarray}
P(t,b_e(t))-P_e(t,b_e(t))&=&P(t,x)-(K-b_e(t))\nonumber\\
&\geq& P(t,b(t))+(b_e(t)-b(t))\partial_x^+P(t,b(t))-(K-b_e(t)) \nonumber \\
&=& (b_e(t)-b(t))(\partial_x^+P(t,b(t))+1).\label{eq-vitess-b1}
\end{eqnarray}
We now use the following lower bound for the jump of derivative of
$P(t,\cdot)$ at $b(t)$ (see \cite{LambertonMikou}, Remark 4.1).
\[
\partial_x^+P(t,b(t))+1\geq \frac{d^+}d,
\]
with $d=d^++\int (e^y-1)_-\nu(dy)$.
By combining  \eqref{eq-vitess-b0} and \eqref{eq-vitess-b1}, we get
\begin{eqnarray*}
 0\leq b_e(t)-b(t)\leq \frac{rK d}{d^+}
      \E\left(\int_0^{T-t}\mathbf{1}_{\{S_s^{b_e(t)}\leq b(t+s)\}}ds\right).
\end{eqnarray*}
We  now want to prove that
\begin{eqnarray}\label{eq-cp-cpe-1}
\lim_{t\to T}\frac 1 {T-t}\E\left(\int_0^{T-t}\mathbf{1}_{\{S_s^{b_e(t)}\leq
       b(t+s)\}}ds\right)=0.
\end{eqnarray}
 We first note that
$$
\E\left(\int_0^{T-t}\mathbf{1}_{\{S_s^{b_e(t)}\leq b(t+s)\}}ds\right)
=\int_0^{T-t}\P\left((r-\delta)s+X_s \leq
       \ln\left(\frac{b(t+s)}{b_e(t)}\right)\right)ds.
$$
Using the notation $\theta=T-t$ and $\zeta(u)=\frac{K}{b_e(T-u)}$,
for $u\in (0,T]$, we have
\begin{eqnarray*}
\ln\left(\frac{b(t+s)}{b_e(t)}\right)
        &\leq& \ln\left(\frac{b_e(t+s)}{b_e(t)}\right)\\
        &\leq& \frac{b_e(t+s)}{b_e(t)}-1\\
        &=&\frac{\zeta(\theta)+1}{\zeta(\theta-s)+1}-1\\
        &=&\frac{\zeta(\theta)-\zeta(\theta-s)}{\zeta(\theta-s)+1}\\
        &\leq&\left| \zeta(\theta)-\zeta(\theta-s)\right|,
\end{eqnarray*}
since $\zeta\geq0$. Therefore,
\begin{eqnarray}\label{eq-vcp-ieq1}
\E\left(\int_0^{\theta}\mathbf{1}_{\{S_s^{b_e(t)}< b(t+s)\}}ds\right)\leq
\int_0^{\theta}\P\left((r-\delta)s+X_s \leq
       |\zeta(\theta)-\zeta(\theta-s)|\right)ds.
\end{eqnarray}
It follows  from Theorem~\ref{thm-vpce-vf} that
\[
\lim_{u\to0}\frac{\zeta(u)}u=\int(e^y-1)_-\nu(dy).
\]
Therefore, given any $\varepsilon>0$, there exists
$\eta_\epsilon>0$ such that,
for $u\in (0,\eta_\epsilon]$,
$$
-\varepsilon+\int(e^y-1)_-\nu(dy)\leq\frac{\zeta(u)}u\leq
\varepsilon+\int(e^y-1)_-\nu(dy).
$$
Take  $\theta\in]0,\eta]$ and $s\in]0,\theta]$. We have
\begin{eqnarray*}
\zeta(\theta)-\zeta(\theta-s)
        &\leq& \theta\left(\varepsilon+\int(e^y-1)_-\nu(dy)\right)-
           (\theta-s)\left(-\varepsilon+\int(e^y-1)_-\nu(dy)\right)\\
        &=& s\int(e^y-1)_-\nu(dy)+2\theta\varepsilon-s\varepsilon\\
        &\leq&s\int(e^y-1)_-\nu(dy)+2\theta\varepsilon.
\end{eqnarray*}

Hence, using \eqref{eq-decom-X-vf} and \eqref{eq-cd-martin-vf}, we get,
with the notation $Z_s=X_s-\gamma_0 s$,
 \begin{eqnarray}
\P\left((r-\delta)s+X_s \leq \left|\zeta(\theta)-\zeta(\theta-s)\right|\right)
        &\leq& \P\left((r-\delta)s+X_s \leq s\int(e^y-1)_-\nu(dy)+2\theta\varepsilon\right)\nonumber\\
        \nonumber
        &=& \P\left(Z_s \leq -s\left(r-\delta+\gamma_0-\int(e^y-1)_-\nu(dy)\right)
              +2\theta\varepsilon\right)\\
        &=& \P\left(Z_s \leq -sd^++2\theta\varepsilon\right).
        \label{eq-vpc-ieq2}
\end{eqnarray}
Now, take
 $\varepsilon<\frac{d^+}{4}$ and $\theta\leq \eta_\epsilon$.
 We deduce
from \eqref{eq-vcp-ieq1} and \eqref{eq-vpc-ieq2}  that
\begin{eqnarray*}
\E\left(\int_0^{\theta}\mathbf{1}_{\{S_s^{b_e(t)}\leq b(t+s)\}}ds\right)
        &\leq& \int_0^{\theta}\P\left(Z_s \leq -sd^++2\theta\varepsilon\right)ds\\
        &=& \int_0^{\frac{4\theta\varepsilon}{d^+}}\P\left(Z_s \leq -sd^++2\theta\varepsilon\right)ds
        +\int_{\frac{4\theta\varepsilon}{d^+}}^{\theta}\P\left(Z_s \leq -sd^++2\theta\varepsilon\right)ds\\
        &\leq& \frac{4\theta\varepsilon}{d^+} +\int_{\frac{4\theta\varepsilon}{d^+}}^{\theta}\P\left(\frac{Z_s}s
        \leq -d^++\frac{2\theta\varepsilon}{s}\right)ds\\
        &\leq& \frac{4\theta\varepsilon}{d^+} +\int_{\frac{4\theta\varepsilon}{d^+}}^{\theta}\P\left(\frac{Z_s}s \leq -\frac{d^+}2\right)ds\\
        &\leq& \frac{4\theta\varepsilon}{d^+} +\int_0^{\theta}\P\left(\frac{Z_s}s \leq -\frac{d^+}2\right)ds.\\
\end{eqnarray*}
It follows from
  Lemma~\ref{YV0} that
    $\displaystyle\lim_{s\to 0}\frac{Z_s}{s}=0$ a.s., so that
 $$\lim_{s\to0}\P\left(\frac{Z_s}s \leq -\frac{d^+}2\right)=0.$$
 Hence
 \[
 \limsup_{\theta\downarrow 0}\frac{1}{\theta}\E\left(
     \int_0^{\theta}\mathbf{1}_{\{S_s^{b_e(t)}\leq b(t+s)\}}ds\right)
     \leq \frac{4\varepsilon}{d^+}.
 \]
Since $\epsilon$ can be arbitrarily close to 0, \eqref{eq-cp-cpe-1} is proved.
\end{proof}

\section{The critical price near maturity in an infinite variation Lévy model}

Throughout this section, we assue that  $X$ is an infinite variation Lévy process i.e.
\[
\sigma\neq0\quad\text{or}\quad\int_{|x|\leq1}|x|\nu(dx)=\infty.
\]
Our main result is that, in this case, the  convergence of $b(t)$ to $K$ cannot be linear.
\begin{thm}\label{thm-infiniteV}
Assume that $X$ is L\'evy process with infinite variation.
If $d^+\geq0$, we have
$$\lim_{t\to T}\frac{1}{T-t} \left(\frac K{b(t)}-1\right)=\infty.$$
\end{thm}
This result follows from the following Lemma, which will be proved later.
\begin{lem}\label{lem-levy-vi}
If  $X$ is a L\'evy process with infinite variation,
 we have
$$\lim_{t\to0}\E\left(\frac{X_t}t\right)_+=\infty.$$
\end{lem}
\noindent {\bf Proof of Theorem~\ref{thm-infiniteV}:}
We use the notation $\theta=T-t$. From the equality $P_e(t,b_e(t))=K-b_e(t)$,
we derive, as in the proof of Theorem~\ref{thm-vpce-vf} (see \eqref{5.1*}), that
\[
(r-\delta)\theta= \E\left[\left(e^{(r-\delta)\theta+X_\theta}-\frac{K}{b_e(t)}\right)_+\right]+o(\theta)
\]
Denote  $\zeta(\theta)=\frac{K}{b_e(t)}-1$. Using the inequality $e^x\geq x+1$, we deduce
\begin{eqnarray*}
(r-\delta)\theta&=& \E\left[\left(e^{(r-\delta)\theta+X_\theta}-1-\zeta(\theta)\right)_+\right]+o(\theta)\\
&\geq& \E\left(\left((r-\delta)\theta+X_\theta-\zeta(\theta)\right)_+\right)+o(\theta)\\
&\geq& \E\left[(r-\delta)\theta+X_\theta\right]_+-\zeta(\theta)+o(\theta)\\
&=& \E\left(\tilde {X}_\theta\right)_+-\zeta(\theta)+o(\theta),
\end{eqnarray*}
where $\tilde{X}_t=(r-\delta)t+X_t$. Therefore,
\begin{eqnarray*}
\liminf_{\theta\downarrow0}\frac{\zeta(\theta)}\theta \geq \lim_{\theta\downarrow 0}
          \E\left(\frac{\tilde{X}_\theta}\theta\right)_+- (r-\delta).
\end{eqnarray*}
Since $\tilde X$ is a L\'evy process with infinite variation,
the Theorem follows from Lemma~\ref{lem-levy-vi}.
\qed

\noindent{\bf Proof of Lemma~\ref{lem-levy-vi}: }
Denote by $(\sigma^2,\gamma,\nu)$ the characteristic triplet  of  $X$.
The L\'evy-It\^o decomposition of $X$ can be written (see \eqref{eq-decom-X} and \eqref{Y})
\[
X_t=\gamma_t+\sigma B_t+\hat{X}_t+X^0_t,\quad t\geq 0,
\]
with
\[
\hat{X}_t=\int_0^t\int_{\{|x|>{1}\}}{x}J_{X}(ds,dx) \quad\mbox{and}\quad
  \tilde{X}^0_t = \int_0^t \int_{\{0<{|x|}\leq{1}\}}{x}\tilde{J}_{X}(ds,{dx}),
\]
where $J_X$ is the jump measure of $X$. Note that $\hat{X}$ is a compound Poisson process.
We  have
\begin{eqnarray*}
\left(X_t\right)_+&=&\left(\gamma_t+\sigma B_t+\tilde{X}^0_t+\hat{X}_t\right)
              \mathbf{1}_{\left\{\gamma_t+\sigma B_t+\tilde{X}^0_t+\hat{X}_t\geq 0\right\}}\\
&\geq&\left(\gamma t + \sigma B_t+\tilde{X}^0_t\right)\mathbf{1}_{\left\{
           \gamma t + \sigma B_t+ \tilde{X}^0_t\geq0\right\}}\mathbf{1}_{\left\{\hat X_t=0\right\}}.
\end{eqnarray*}
Since $B$, $\hat X$ and $\tilde{X}^0$ are independent,
we have
\begin{eqnarray*}
\E\left(\frac{X_t}t\right)_+
&\geq& \E\left(\frac{\sigma B_t+ \tilde{X}^0_t}t+\gamma\right)_+\P(\hat X_t=0)\\
&\geq& \E\left(\frac{\sigma B_t+ \tilde{X}^0_t}t+\gamma\right)_+e^{-t\nu(\{|x|\geq1\})},
\end{eqnarray*}
where the last inequality follows from the fact that the first jump time of the process $\hat X$
is exponentially distributed with parameter $\nu(\{|x|\geq1\})$.
Since
$\left(\frac{\sigma B_t+ \bar X^0_t}t+\gamma\right)_+\geq
     \left(\frac{\sigma B_t+ \bar X^0_t}t\right)_+-|\gamma|,$
 it suffices to show that
\begin{eqnarray}\label{eq-lem-vi}
\lim_{t\to0}\E\left[\left(\frac{\sigma B_t+ \tilde X^0_t}t\right)_+\right]=\infty.
\end{eqnarray}
For that, we discuss two cases.

We first assume that $\sigma\neq 0$.
Recall that $B$ and $\tilde X^0$ are independent and $\E(\bar X^0_t)=0$.
By conditioning  on $B$ and using Jensen's inequality, we get
\begin{eqnarray*}
\E\left(\frac{\sigma B_t+ \tilde X^0_t}t\right)_+
&=& \E\left[\E\left(
            \left(\frac{\sigma B_t+ \tilde X^0_t}t\right)_+|B_t\right)\right]\\
&\geq& \E\left[\left(
         \E\left(\frac{\sigma B_t+ \tilde X^0_t}t|B_t\right)\right)_+\right]\\
&=& \E\left[\left(\sigma\frac{B_t}t\right)_+\right]\\
&=& |\sigma|\frac{1}{\sqrt{2\pi t}},
\end{eqnarray*}
so that  \eqref{eq-lem-vi} is proved.

Now, assume $\sigma=0$. Since the process $X$ has infinite variation,
we must have
\[
\displaystyle\int_{|y|\leq1}|y|\nu(dy)=\infty.
\]
Given $\epsilon\in (0,1)$ and $t>0$, introduce
\begin{eqnarray*}
 \tilde X^\epsilon_t&=&
      \int_0^t \int_{\{\epsilon\leq{|x|}\leq{1}\}}{x}\tilde{J}_{X}(ds,{dx})\\
      &=&X^\epsilon_t-C_\epsilon t,
\end{eqnarray*}
where
\[
X^\varepsilon_t=\sum_{0\leq s\leq t}
                 \Delta X_s\ind{\{\varepsilon\leq|\Delta X_s|\leq1\}}
                      \quad\mbox{and}\quad C_\epsilon=
                                       \int_{\{\epsilon\leq|y|\leq1\}}y\nu(dy).
\]
We have
$\tilde X^0_t=\tilde X^\varepsilon_t+(\tilde X^0_t-\tilde X^\varepsilon_t)$,
and the random variables
 $\tilde X^\varepsilon_t$ and $\tilde X^0_t-\tilde X^\varepsilon_t$
are independent and centered. Therefore
\begin{eqnarray}
\E\left[\left(\frac{\tilde X^0_t}t\right)_+\right]
&\geq&\E\left[\left(\frac{\tilde{X}^{\varepsilon}_t}t\right)_+\right]\nonumber\\
&=&\E\left(\frac{X^\varepsilon_t-tC_\varepsilon}{t}\right)_+
         \label{eq-lem-vi-1}.
\end{eqnarray}
We have $\E\left(X^\varepsilon_t-tC_\varepsilon\right)_+=
   \E g_t(X^\epsilon_t)$, with $g_t(x)=(x-tC_\varepsilon)_+$.
   Since $X^\varepsilon$
is a compound Poisson process,
\[
g_t(X^\varepsilon_t)-g_t(0)=\sum_{0 < s\leq t}[g_t(X^\varepsilon_s)-g_t(X^\varepsilon_{s^-})],
\]
so that, due to the compensation formula
(cf. Proposition~\ref{prop-formule-comp}),
\begin{eqnarray*}
\E\left(X^\varepsilon_t-tC_\varepsilon\right)_+&=&g_t(0)
    +
\E\left(\sum_{0\leq s\leq t}\left(g_t(X^\varepsilon_s)-g_t(X^\varepsilon_{s^-})
        \right)\right)\\
        &=& t\left(-C_\epsilon\right)_++\E\left(\int_0^t ds
        \int_{\{\epsilon\leq |y|\leq 1\}}
        \left(g_t(X^\varepsilon_s+y)-
            g_t(X^\varepsilon_s)\right)\nu(dy)\right).
\end{eqnarray*}
For any fixed $\epsilon\in(0,1)$, we have
\begin{eqnarray*}
\E\left(\int_0^t ds\int_{\{\epsilon\leq |y|\leq 1\}}
        \left(g_t(X^\varepsilon_s+y)-
            g_t(X^\varepsilon_s)\right)\nu(dy)\right)
        &=& \E\left[\int_0^t ds\int\left[(X^\varepsilon_s+y)_+
             -(X^\varepsilon_s)_+\right]\nu(dy)\right]+o(t)\\
        &=& t\int_{\{\epsilon\leq |y|\leq 1\}} y_+\nu(dy)+o(t).
        \end{eqnarray*}
Therefore,
\begin{eqnarray*}
\E[(X^\varepsilon_t-tC_\varepsilon)_+]
&=& t\left[(- C_\varepsilon)_+ + \int_{\{\epsilon\leq |y|\leq 1\}} y_+\nu(dy)\right]+o(t).
\end{eqnarray*}
Going back to  \eqref{eq-lem-vi-1}, we derive
\begin{eqnarray*}
\liminf_{t\to 0}\E\left[\left(\frac{\tilde X^0_t}t\right)_+\right]
&\geq& (-C_\varepsilon)_+ +\int_{\{\epsilon\leq |y|\leq 1\}}y_+\nu(dy)\\
&=&\left(
      -\int_{\{\epsilon\leq |y|\leq 1\}} y\nu(dy)\right)_+
           +\int_{\{\epsilon\leq |y|\leq 1\}}y_+\nu(dy).
\end{eqnarray*}
Since $\int_{\{|y|\leq1\}} |y|\nu(dy)=\infty$, we have either
$\lim_{\epsilon\downarrow 0}\int_{\{\epsilon \leq|y|\leq1\}}y_+\nu(dy)=\infty$
or
\[
\lim_{\epsilon\downarrow 0}\left(\int_{\{\epsilon \leq |y|\leq1\}}y\nu(dy)
\right)_+=\infty,
\]
and \eqref{eq-lem-vi}
follows.
\qed
\section{Critical price and tempered stable processes}\label{sec-stable}
Throughout this section,  the following assumption is in force.
 \begin{description}
 \item[(AS)] We have
  \[
  \E\left(e^{iu X_t}\right)=\exp\left(
                t\int\left(e^{iuy}-1-iu(e^y-1)\right)\nu(dy)\right),
  \]
  with $\int (e^y-1)_+\nu(dy)<r-\delta$ and, for some $a_0<0$,
  \[
  \ind{\{a_0<y<0\}}\nu(dy)=\frac{\eta(y)}{|y|^{1+\alpha}}
                                          \ind{\{a_0<y<0\}}dy,
  \]
  where $1<\alpha <2$ and $\eta$ is a positive bounded Borel
  measurable function on $[a_0,0)$, which satisfies
   $\lim_{y\to 0}\eta (y)=\eta_0>0$.
 \end{description}
Note that, under this assumption, we have $\nu[(-\infty,0)]>0$, so that
\eqref{eq-nonsub} is satisfied.

\begin{thm}\label{thm-Stable}
Under assumption (AS), we have
\[
\lim_{t\to T}
     \frac{K-b(t)}{(T-t)^{1/\alpha}|\ln( T-t)|^{1-\frac{1}{\alpha}}}
     =K\left( \eta_0\frac{
               \Gamma(2-\alpha)}{\alpha-1}\right)^{1/\alpha}.
\]
\end{thm}
For the proof of Theorem~\ref{thm-Stable}, we use the same approach
as in Section~\ref{SectionFV}. Namely, we first characterize the rate of
convergence of the {\em European} critical price $b_e(t)$: this is done
in Section~\ref{StableEurope} (see Proposition\ref{prop-ASe}
and recall that $b(t)\leq b_e(t)\leq K$). Then,
we estimate the difference between the European and the American
critical prices. In fact, Theorem~\ref{thm-Stable} is a direct consequence
of Proposition\ref{prop-ASe}, combined with
Proposition~\ref{Prop-StableAmEu}.


Before investigating the behavior of the European critical price, we establish
a crucial consequence of assumption (AS), namely the fact that, for small
$t$, the L\'evy process at time $t$ behaves asymptotically like a
one sided stable  random variable of order $\alpha$.
 \begin{lem}\label{lem-AS1}Under assumption (AS),
as $t$ goes to $0$, the random variable $X_t/t^{1/\alpha}$ converges in
distribution to a random variable $Z$ with characteristic function given by
\[
\E\left(e^{iu Z}\right)=\exp\left(
      \eta_0\int_0^{+\infty}\left(e^{-iuz}-1+iuz\right)\frac{dz}{z^{1+\alpha}}\right),
      \quad u\in\R.
\]
\end{lem}
\begin{pf}
Introduce the following decomposition of the process $X$
\[
X_t=X^0_t-t \int (e^y-1)_+\nu(dy)+\bar{X}_t, \quad t\geq 0,
\]
where
\[
X^0_t=\sum_{0< s\leq t}\Delta X_s \ind{\{\Delta X_s>0\}}.
\]
Note that the process $X^0$ is well defined because
$\int y_+\nu(dy)\leq \int (e^y-1)_+\nu(dy)<\infty$,
and the characteristic function of $\bar{X}_t$ is given by
\begin{equation}\label{eq-carac}
\E\left(e^{iu\bar{X}_t}\right)=\exp\left(
                 t\int_{(-\infty,0)}\left(e^{iuy}-1-iu(e^y-1)\right)\nu(dy)
                                        \right), \quad u\in \R.
\end{equation}

We have $\lim_{t\downarrow 0}\frac{X^0_t}{t}=0$ a.s.
(see \cite{sato}, Section 47), so that, with probability one,
\[
\lim_{t\downarrow 0}\frac{X_t-\bar{X}_t}{t^{1/\alpha}}=0.
\]
We will now prove that $\bar{X}_t/t^{1/\alpha}$ weakly converges to $Z$
as $t\to 0$. For a fixed $u\in \R$, we have
\[
\E\left(e^{iu\frac{\bar{X}_t}{t^{1/\alpha}}}\right)=\exp\left(
                 t\int_{(-\infty,0)}\left(e^{iuy/t^{1/\alpha}}-1
                 -
                \frac{iu}{t^{1/\alpha}}(e^y-1)\right)\nu(dy)
                                        \right).
\]
The integral in the exponential can be split in two parts
\begin{eqnarray*}
\int_{(-\infty,0)}\left(e^{iuy/t^{1/\alpha}}-1
                 -
                \frac{iu}{t^{1/\alpha}}(e^y-1)\right)\nu(dy)
                &=&\int_{(-\infty,a_0]}\left(e^{iuy/t^{1/\alpha}}-1
                 -
                \frac{iu}{t^{1/\alpha}}(e^y-1)\right)\nu(dy)\\
               && +
                \int_{a_0}^0\left(e^{iuy/t^{1/\alpha}}-1
                 -
                \frac{iu}{t^{1/\alpha}}(e^y-1)\right)\nu(dy).
\end{eqnarray*}
We have
\begin{eqnarray*}
\left|\int_{(-\infty,a_0]}\left(e^{iuy/t^{1/\alpha}}-1
                 -
                \frac{iu}{t^{1/\alpha}}(e^y-1)\right)\nu(dy)
                \right|
                &\leq &
                2\nu((-\infty,a_0])+\frac{|u|}{t^{1/\alpha}}\int_{(-\infty,a_0]}
                     |e^y-1|\nu(dy),
\end{eqnarray*}
so that
\[
\lim_{t\downarrow 0}\left(t\int_{(-\infty,a_0]}\left(e^{iuy/t^{1/\alpha}}-1
                 -
                \frac{iu}{t^{1/\alpha}}(e^y-1)\right)\nu(dy)\right)=0.
\]
On the other hand,
\begin{eqnarray*}
\int_{a_0}^0\left(e^{iuy/t^{1/\alpha}}-1
                 -
                \frac{iu}{t^{1/\alpha}}(e^y-1)\right)\nu(dy)
                &=&
                \int_{a_0}^0\left(e^{iuy/t^{1/\alpha}}-1
                 -
                \frac{iu}{t^{1/\alpha}}y\right)\frac{\eta(y)}{|y|^{1+\alpha}}dy\\
               && +\frac{iu}{t^{1/\alpha}}
                \int_{a_0}^0\left(y-(e^y-1)
                \right)\frac{\eta(y)}{|y|^{1+\alpha}}dy\\
                &=&
                \int_{0}^{|a_0|}\left(e^{-iuy/t^{1/\alpha}}-1
                 +
                \frac{iu}{t^{1/\alpha}}y\right)\frac{\eta(-y)}{y^{1+\alpha}}dy
                +O(t^{-1/\alpha}),
\end{eqnarray*}
where the last equality follows from the boundedness of $\eta$
and the fact that $\int_{(a_0,0)} y^2\nu(dy)<\infty$.
Hence, using the substitution $z=y/t^{1/\alpha}$,
\begin{eqnarray*}
\int_{a_0}^0\left(e^{iuy/t^{1/\alpha}}-1
                 -
                \frac{iu}{t^{1/\alpha}}(e^y-1)\right)\nu(dy)
                &=&\frac{1}{t}
  \int_{0}^{\frac{|a_0|}{t^{1/\alpha}}}\left(e^{-iuz}-1
                 +
               iuz\right)\frac{\eta(-z/t^{1/\alpha})}{z^{1+\alpha}}dz
                +O(t^{-1/\alpha}),
\end{eqnarray*}
so that, by dominated convergence,
\[
\lim_{t\downarrow 0}\left(
    t\int_{a_0}^0\left(e^{iuy/t^{1/\alpha}}-1
                 -
                \frac{iu}{t^{1/\alpha}}(e^y-1)\right)\nu(dy)
                \right)=
           \int_{0}^{+\infty}\left(e^{-iuz}-1
                 +
               iuz\right)\frac{\eta_0}{z^{1+\alpha}}dz,
\]
and the lemma is proved.
\end{pf}
\subsection{European critical price}\label{StableEurope}
Denote $\theta =T-t$. The equality $P_e(t,b_e(t))=K-b_e(t)$ can be written as follows
\begin{eqnarray*}
K-b_e(t)&=&\E e^{-r\theta}\left(K-b_e(t)e^{(r-\delta)\theta+X_\theta}\right)_+\\
  &=&Ke^{-r\theta}-b_e(t)e^{-\delta\theta}+
                            \E e^{-r\theta}\left(b_e(t)e^{(r-\delta)\theta+X_\theta}-K\right)_+.
\end{eqnarray*}
 Hence
    \ben\label{**}
    \frac{K}{b_e(t)}\left(1-e^{-r\theta}\right)-\left(1-e^{-\delta\theta}\right)=\E e^{-r\theta}\left(e^{(r-\delta)\theta+X_\theta}-\frac{K}{b_e(t)}\right)_+.
    \een
    Since $\lim_{t\to T} b_e(t)=K$, the left-hand side is equal to $(r-\delta)\theta+o(\theta)$. For the study of the right-hand side, let
    \[
    \zeta(\theta)=\frac{K}{b_e(t)}-1,
        \]
 so that from (\ref{**}) we derive
 \begin{eqnarray}
(r-\delta)\theta&=&\E e^{-r\theta}\left(e^{(r-\delta)\theta+X_\theta}
                                         -1-\zeta(\theta)\right)_+ +o(\theta)\nonumber\\
                        &=&\E \left(e^{(r-\delta)\theta+X_\theta}
                                         -1-\zeta(\theta)\right)_+ +o(\theta)\label{**'},
                              \end{eqnarray}
where we have used the fact that
$\lim_{\theta\to 0}\E\left(e^{(r-\delta)\theta+X_\theta}
                                         -1-\zeta(\theta)\right)_+=0$.
The following statement clarifies the behavior of $\zeta(\theta)$ as
$\theta\downarrow 0$.
\begin{prop}\label{prop-ASe}
Under assumption (AS), we have
\[
\lim_{\theta\downarrow 0}
     \frac{\zeta(\theta)}{\theta^{1/\alpha}|\ln \theta|^{1-\frac{1}{\alpha}}}
     =\left(\alpha \eta_0I_\alpha\right)^{1/\alpha}
               =\left( \eta_0\frac{
               \Gamma(2-\alpha)}{\alpha-1}\right)^{1/\alpha},
\]
where
\[
I_\alpha =\int_0^{+\infty}\left(e^{-z}-1+z\right)\frac{dz}{z^{1+\alpha}}
     =\frac{\Gamma(2-\alpha)}{\alpha(\alpha-1)}.
\]
\end{prop}
The first step in the proof of Proposition~\ref{prop-ASe}
is the following lemma.
\begin{lem}\label{lem-AS2}
  We have
    \ben\label{eq-AS2.1}
  \lim_{\theta\to 0}\frac{\zeta(\theta)}{\theta^{1/\alpha}}=+\infty.
  \een
\end{lem}
\begin{pf}
Note that
\begin{eqnarray*}
\left|\E \left(e^{(r-\delta)\theta+X_\theta}
                                         -1-\zeta(\theta)\right)_+ -
                                 \E \left(e^{X_\theta}
                                         -1-\zeta(\theta)\right)_+
                                         \right|
                                         &\leq&
                                      \left(e^{(r-\delta)\theta}-1\right)
                                      \E\left(e^{X_\theta}\right)
                                      =O(\theta),
\end{eqnarray*}
so that, in view of (\ref{**'}), we have
\[
\E \left(e^{X_\theta}
                                         -1-\zeta(\theta)\right)_+
                                         =
                                         O(\theta).
\]
Since $e^x\geq 1+x$, we also have
$\E\left( X_\theta-\zeta(\theta)\right)_+=O(\theta)$.
Therefore
\[
\lim_{\theta \downarrow 0}\E\left(\frac{X_\theta}{\theta^{1/\alpha}}-
      \frac{\zeta(\theta)}{\theta^{1/\alpha}}\right)_+=0.
\]
If we had
$\liminf_{\theta\downarrow 0}\zeta(\theta)/\theta^{1/\alpha}=
\lambda\in[0,+\infty)$,
we would deduce from Lemma~\ref{lem-AS1} and Fatou's Lemma that
\[
\E\left(Z-\lambda\right)_+=0.
\]
Hence $\P(Z\leq \lambda)=1$. However, the support of the random variable
$Z$ (which is a one-sided stable random variable of order $\alpha$) is
the whole real line. This proves (\ref{eq-AS2.1}) by contradiction.
\end{pf}

The next lemma provides some estimates for the moment generating
function of the process $\bar{X}$, defined by
\[
\bar{X}_t=X_t+t \int (e^y-1)_+\nu(dy)-\sum_{0<s\leq t}
            \Delta X_s\ind{\{\Delta X_s>0\}}, \quad t\geq 0.
\]
\begin{lem}\label{lem-AS3}
We have, for all $\rho\geq 0$, $t\geq 0$,
\[
\E\left(e^{\rho \bar{X}_t}\right)=e^{t\bar{\phi}(\rho)},
\]
with
\begin{equation}\label{eq-barphi}
\bar{\phi}(\rho)=\int_{(-\infty,0)}\left(e^{\rho y}-1-
         \rho(e^y-1)\right)\nu(dy), \quad \rho\geq 0.
\end{equation}
Morover, for any $a\in [a_0,0)$ and any $\rho\geq 0$, we have
\[
\rho^\alpha H_a(\rho)-\nu_a-\rho\bar{\nu}_a\leq \bar{\phi}(\rho)\leq
         \rho \nu_a+\rho^\alpha H_a(\rho),
\]
where
\[
\nu_a=\nu((-\infty,a]),\quad
\bar{\nu}_a=\int_0^{|a|}\left(e^{-y}-1+
         y\right)\nu(dy)\quad
           \mbox{and}\quad
H_a(\rho)=  \int_0^{|a|\rho}
      \left(e^{-z}-1+
         z)\right)\frac{\eta(-z/\rho)}{z^{1+\alpha}}dz.
     \]
\end{lem}
\begin{pf}
First, note that \eqref{eq-barphi} is deduced from \eqref{eq-carac}
by analytic continuation.
Now, fix $\rho\geq 0$ and $a\in [a_0,0)$.
We have,
\begin{eqnarray*}
\bar{\phi}(\rho)&=&\int_{(-\infty,a]}\left(e^{\rho y}-1-
         \rho(e^y-1)\right)\nu(dy)+\bar{\phi}_a(\rho)
  \end{eqnarray*}
  with the notation
  \[
\bar{\phi}_a(\rho)=\int_a^0\left(e^{\rho y}-1-
         \rho(e^y-1)\right)\nu(dy),\quad \rho\geq 0.
  \]
  For $y\in (-\infty,a]$, we have
  $-1\leq e^{\rho y}-1-
         \rho(e^y-1)\leq \rho$. Therefore
\[
\bar{\phi}_a(\rho)-\nu_a\leq \bar{\phi}(\rho)\leq \bar{\phi}_a(\rho)+\rho\nu_a.
\]
On the other hand,
\begin{eqnarray*}
\bar{\phi}_a(\rho)&=&\int_0^{|a|}
      \left(e^{-\rho y}-1-
         \rho(e^{-y}-1)\right)\frac{\eta(-y)}{y^{1+\alpha}}dy\\
         &=&
         \int_0^{|a|}
      \left(e^{-\rho y}-1+
         \rho y)\right)\frac{\eta(-y)}{y^{1+\alpha}}dy
         -\rho\int_0^ {|a|}
      \left(
         e^{-y}-1+y\right)\frac{\eta(-y)}{y^{1+\alpha}}dy.
\end{eqnarray*}
We have
$ e^{-y}-1+y\geq 0$. Hence
\[
   -\rho \int_0^ {|a|}
      \left(
         e^{-y}-1+y\right)\frac{\eta(-y)}{y^{1+\alpha}}dy
         +\psi_a(\rho)\leq\bar{\phi}_a(\rho)\leq
         \psi_a(\rho),
\]
where
\begin{eqnarray*}
\psi_a(\rho)
=
         \int_0^{|a|}
      \left(e^{-\rho y}-1+
         \rho y)\right)\frac{\eta(-y)}{y^{1+\alpha}}dy=
                  \rho^\alpha
         \int_0^{|a|\rho}
      \left(e^{-z}-1+
         z)\right)\frac{\eta(-z/\rho)}{z^{1+\alpha}}dz=\rho^\alpha H_a(\rho).
\end{eqnarray*}
\end{pf}

The crucial step in the proof of Proposition~\ref{prop-ASe}
is an asymptotic estimate for the tail of the distribution
of $\bar{X}_\theta/\theta^{1/\alpha}$ as $\theta$ approaches $0$.
This will be given in Lemma~\ref{lem-asymp}. We first give a
preliminary uniform bound.
\begin{lem}\label{lem-tail}
Let $a\in [a_0,0)$. There exists a positive constant $C_a$ such
that, for all $\theta >0$, $t>0$, we have
\[
\ln \P\left(\frac{\bar{X}_\theta}{\theta^{1/\alpha}}\geq t\right)
  \leq C_a\theta^{1-\frac{1}{\alpha}} t^{\frac{1}{\alpha-1}}
       -J_\alpha(a)t^{\frac{\alpha}{\alpha-1}},
\]
where
\[
J_\alpha(a)=\frac{\alpha-1}{\alpha^{\frac{\alpha}{\alpha-1}}\left(
          \eta^*(a)I_\alpha\right)^{\frac{1}{\alpha-1}}},
\]
with
\[
\eta^*(a)=\sup_{u\in (a,0)}\eta(u)\quad\mbox{and}\quad
I_\alpha=
 \int_0^{+\infty}\left(e^{-z}-1+z\right)\frac{dz}{z^{1+\alpha}}
  =\frac{\Gamma(2-\alpha)}{\alpha(\alpha-1)}.
\]
\end{lem}
\begin{pf}
For any $p>0$, we have, using Markov's inequality and
Lemma~\ref{lem-AS3},
\begin{eqnarray*}
\P\left(\frac{\bar{X}_\theta}{\theta^{1/\alpha}}\geq t\right)&\leq&
          e^{-pt}\E\left(e^{p\bar{X}_\theta/\theta^{1/\alpha}}\right)\\
          &=&
          e^{-pt}e^{\theta\bar{\phi}\left(p/\theta^{1/\alpha}\right)}\\
          &\leq &
          e^{-pt}e^{p^\alpha H_a(p/\theta^{1/\alpha})+ \theta^{1-\frac{1}{\alpha}}p\nu_a}\\
          &\leq &
          e^{-pt}e^{p^\alpha\eta^*(a)I_\alpha+
             \theta^{1-\frac{1}{\alpha}}p\nu_a},
\end{eqnarray*}
where the last inequality follows from $H_a(\rho)\leq \eta^*(a)I_\alpha$.
By choosing $p=\left(\frac{t}{\alpha \eta^*(a)I_\alpha}\right)^{1/(\alpha-1)}$,
we get
\begin{eqnarray*}
\P\left(\frac{\bar{X}_\theta}{\theta^{1/\alpha}}\geq t\right)&\leq&
    \exp\left(-J_\alpha(a) t^{\frac{\alpha}{\alpha-1}}+
    C_a\theta^{1-\frac{1}{\alpha}} t^{\frac{1}{\alpha-1}}\right),
\end{eqnarray*}
with $C_a=\left(\alpha\eta^*(a)I_\alpha\right)^{-\frac{1}{\alpha-1}}\nu_a$.
\end{pf}

We are now in a position to prove the main estimate
for the proof of Proposition~\ref{prop-ASe}.
\begin{lem}\label{lem-asymp}
Denote, for $\theta>0$,
$\displaystyle
\bar{Z}_\theta =\frac{\bar{X}_\theta}{\theta^{1/\alpha}}$.
We have, for any function
$\xi: (0,+\infty)\to (0,+\infty)$ satisfying
$\displaystyle
\lim_{\theta\downarrow 0}\xi(\theta)=+\infty$,
\[
\lim_{\theta\downarrow 0}\frac{\ln \P(\bar{Z}_\theta\geq \xi(\theta))}{
       (\xi(\theta))^{\frac{\alpha}{\alpha-1}}}=-J_\alpha(0),
\]
where
\[
J_\alpha(0)=\lim_{a\uparrow 0}J_\alpha(a)=\frac{\alpha-1}{(\alpha^{\alpha}
     \eta_0I_\alpha)^{\frac{1}{\alpha-1}}}.
\]
\end{lem}
\begin{pf}
We first prove
\begin{equation}\label{eq-limsup-tail}
\limsup_{\theta\downarrow 0}
      \frac{\ln \P(\bar{Z}_\theta\geq \xi(\theta))}{
       (\xi(\theta))^{\frac{\alpha}{\alpha-1}}}\leq-J_\alpha(0).
\end{equation}
Applying Lemma~\ref{lem-tail} with $t=\xi(\theta)$, we
have, for all $a\in [a_0, 0)$,
\[
\ln \P(\bar{Z}_\theta\geq \xi(\theta))
  \leq
  -J_\alpha(a)( \xi(\theta))^{\frac{\alpha}{\alpha-1}}
       +C_a\theta^{1-\frac{1}{\alpha}}
       ( \xi(\theta))^{\frac{1}{\alpha-1}}.
\]
Hence
\[
\limsup_{\theta\downarrow 0}
      \frac{\ln \P(\bar{Z}_\theta\geq \xi(\theta))}{
       (\xi(\theta))^{\frac{\alpha}{\alpha-1}}}\leq -J_\alpha(a),
\]
and \eqref{eq-limsup-tail} follows by making $a$ go to $0$.

In order to derive a lower bound for the $\liminf$, we proceed as follows.
Given any $p>0$ and any $t>0$, we have
\begin{eqnarray*}
\E\left(e^{p\bar{Z}_\theta}\right)&=&
 \E\left(e^{p\bar{Z}_\theta}\ind{\{\bar{Z}_\theta<t\}}\right)+
     \E\left(e^{p\bar{Z}_\theta}\ind{\{\bar{Z}_\theta\geq t\}}\right)\\
     &=&
     \E\left(\int_{-\infty}^{\bar{Z}_\theta}pe^{ps}ds
     \ind{\{\bar{Z}_\theta<t\}}\right)
         +\E\left(e^{p\bar{Z}_\theta}\ind{\{\bar{Z}_\theta\geq t\}}\right)\\
         &\leq&
         1+  \E\left(\int_{0}^{+\infty}pe^{ps}
                          \ind{\{0<s\leq\bar{Z}_\theta<t\}}ds\right)
         +\E\left(e^{p\bar{Z}_\theta}\ind{\{\bar{Z}_\theta\geq t\}}\right)\\
         &\leq &
        1+  \int_{0}^{t}pe^{ps}\P\left(\bar{Z}_\theta\geq s\right)ds
                       +\E\left(e^{p\bar{Z}_\theta}\ind{\{\bar{Z}_\theta\geq t\}}\right).
\end{eqnarray*}
It follows from Lemma~\ref{lem-tail} that
$\P\left(\bar{Z}_\theta\geq s\right)\leq
\exp\left(C_a\theta^{1-\frac{1}{\alpha}} s^{\frac{1}{\alpha-1}}
       -J_\alpha(a)s^{\frac{\alpha}{\alpha-1}}\right)$, so that
\begin{equation}\label{eq-liminf1}
\E\left(e^{p\bar{Z}_\theta}\right)
 \leq 1+pF_a(\theta,t)\int_0^t e^{ps-J_\alpha(a)s^{\frac{\alpha}{\alpha-1}}}ds
   +\E\left(e^{p\bar{Z}_\theta}\ind{\{\bar{Z}_\theta\geq t\}}\right),
\end{equation}
with
\[
F_a(\theta, t)=e^{C_a\theta^{1-\frac{1}{\alpha}} t^{\frac{1}{\alpha-1}}}.
\]
For notational convenience, let
\[
\hat{\alpha}=\frac{\alpha}{\alpha-1}, \quad\mbox{so that} \quad
   \hat{\alpha}-1=\frac{1}{\alpha-1},
\]
and
\[
f_p(s)=ps -J_\alpha(a)s^{\hat{\alpha}}, \quad s>0.
\]
We have $f'_p(s)=p-\hat{\alpha}J_\alpha(a)s^{\hat{\alpha}-1}$, so that
the function $f_p$ is increasing on the interval $[0, s^*_p]$
and decreasing on $[s^*_p,+\infty)$, where
\[
s^*_p=\left(\frac{p}{J_\alpha(a)\hat{\alpha}}\right)^{\frac{1}{\hat{\alpha}-1}}
=\left(\frac{p}{J_\alpha(a)\hat{\alpha}}\right)^{\alpha-1}.
\]
We now fix $t>0$ and choose $p=Mt^{\frac{1}{\alpha-1}}$,
where $M$ is a constant satisfying
\[
M\;>\;{J_\alpha(0)\hat{\alpha}}\;=\;\frac{\alpha}{\alpha-1}J_\alpha(0)
    \; =\;\frac{1}{\left(\alpha\eta_0I_\alpha\right)^{\frac{1}{\alpha-1}}}.
\]
We then have $M>J_\alpha(a)\hat{\alpha}$ for all $a\in[a_0,0)$,  so that
\[
t <\left( \frac{M}{J_\alpha(a)\hat{\alpha}}\right)^{\alpha-1}\!t
     \;=\;\left(\frac{p}{J_\alpha(a)\hat{\alpha}}\right)^{\alpha-1}=s^*_p.
\]
Therefore
\[
\forall s\in [0,t],\quad f_p(s)\leq f_p(t)=pt-J_\alpha(a)t^{\hat{\alpha}}=
t^{\hat{\alpha}}\left({M}
                   -J_\alpha(a)\right),
\]
so that
\[
\int_0^t e^{f_p(s)}ds\leq te^{f_p(t)}=t\exp\left(t^{\hat{\alpha}}\left(M
                   -J_\alpha(a)\right)\right).
\]
Going back to \eqref{eq-liminf1}, we get
\begin{eqnarray}
\E\left(e^{Mt^{\frac{1}{\alpha-1}}\bar{Z}_\theta}\right)
 &\leq& 1+Mt^{\frac{1}{\alpha-1}}
         F_a(\theta,t)
         t\exp\left(t^{\hat{\alpha}}\left(M
                   -J_\alpha(a)\right)\right)
   +\E\left(e^{Mt^{\frac{1}{\alpha-1}}\bar{Z}_\theta}
        \ind{\{\bar{Z}_\theta\geq t\}}\right)\nonumber\\
   &=&
   1+Mt^{\hat{\alpha}}
         F_a(\theta,t)
         \exp\left(t^{\hat{\alpha}}\left(M
                   -J_\alpha(a)\right)\right)
   +\E\left(e^{Mt^{\hat{\alpha}-1}\bar{Z}_\theta}
    \ind{\{\bar{Z}_\theta\geq t\}}\right).
   \label{eq-liminf2}
\end{eqnarray}
On the other hand, we have, using Lemma~\ref{lem-AS3},
\begin{eqnarray}
\E\left(e^{Mt^{\frac{1}{\alpha-1}}\bar{Z}_\theta}\right)
   &=&\E\left(e^{Mt^{\hat{\alpha}-1}\bar{X}_\theta/\theta^{1/\alpha}}\right)
             \nonumber\\
   &=&\exp\left(\theta\bar{\phi}(Mt^{\hat{\alpha}-1}/\theta^{1/\alpha})\right)
       \nonumber\\
   &\geq &
     \exp\left[\theta\left(
       \left(\frac{Mt^{\hat{\alpha}-1}}{\theta^{1/\alpha}}\right)^\alpha
          H_a(Mt^{\hat{\alpha}-1}/\theta^{1/\alpha})-\nu_a-
     \frac{Mt^{\hat{\alpha}-1}}{\theta^{1/\alpha}}\bar{\nu}_a\right)\right]
        \nonumber\\
     &=&
     \exp\left( t^{\hat{\alpha}}K_a(M,\theta,t)\right)G_a(M,\theta,t),
     \label{eq-liminf3}
\end{eqnarray}
where
\[
K_a(M,\theta,t)=M^\alpha H_a(Mt^{\hat{\alpha}-1}/\theta^{1/\alpha})
\quad\mbox{and}\quad
G_a(M,\theta,t)=e^{
                 -\theta\nu_a-
                 M \theta^{1-\frac{1}{\alpha}}t^{\hat{\alpha}-1}\bar{\nu}_a}.
\]
Combining \eqref{eq-liminf2} and \eqref{eq-liminf3}, we have
\begin{eqnarray}
\E\left(e^{Mt^{\hat{\alpha}-1}\bar{Z}_\theta}
    \ind{\{\bar{Z}_\theta\geq t\}}\right)&\geq &
    e^{ t^{\hat{\alpha}}K_a(M,\theta,t)}\left(
    G_a(M,\theta,t)-e^{-t^{\hat{\alpha}}K_a(M,\theta,t)}\right.\nonumber\\
    &&\left.-
    Mt^{\hat{\alpha}}
         F_a(\theta,t)
         \exp\left(t^{\hat{\alpha}}\left(M
                   -J_\alpha(a)-K_a(M,\theta,t)\right)\right)
                   \right).\label{eq-liminf4}
\end{eqnarray}
In order to study the sign of $M
                   -J_\alpha(a)-K_a(M,\theta,t)=
                   M-M^\alpha H_a(Mt^{\hat{\alpha}-1}/\theta^{1/\alpha})
                    -J_\alpha(a)$,
we introduce the function
\[
\psi_\alpha(M)=M-M^\alpha\eta_0I_\alpha,\quad M>0.
\]
We have
\begin{eqnarray*}
\psi_\alpha'(M)&=&1-\alpha M^{\alpha-1}\eta_0I_\alpha\\
    &<&0\quad\mbox{ for } M>
      \frac{1}{\left(\alpha\eta_0I_\alpha\right)^{\frac{1}{\alpha-1}}}
             =J_\alpha(0)\hat{\alpha}.
  \end{eqnarray*}
Therefore
\begin{eqnarray*}
\psi_\alpha(M)&<&\psi_\alpha\left(J_\alpha(0)\hat{\alpha}\right)\\
    &=&\psi_\alpha\left(\frac{1}{(\alpha
     \eta_0I_\alpha)^{\frac{1}{\alpha-1}}}\right)\\
     &=&\frac{1}{(\alpha
     \eta_0I_\alpha)^{\frac{1}{\alpha-1}}}\left(1-\frac{1}{\alpha}
     \right)=J_\alpha(0).
\end{eqnarray*}
Since $\lim_{a\uparrow 0}J_\alpha(a)=J_\alpha(0)$, we also have,
for $a$ close to 0,
\[
\psi_\alpha(M)<J_\alpha(a).
\]
Now, consider any function $\xi: (0,+\infty)\to (0,+\infty)$, such that
$
\lim_{\theta\downarrow 0}\xi(\theta)=+\infty$.
 We will apply \eqref{eq-liminf4} with $t=\xi(\theta)$.
Note that $\lim_{\theta\downarrow 0}K_a(M,\theta,\xi(\theta))=
                       M^\alpha\eta_0I_\alpha$,
so that
\[
\lim_{\theta\downarrow 0}\left(
M
                   -J_\alpha(a)-K_a(M,\theta,\xi(\theta))\right)=
                   \psi_\alpha(M)-J_\alpha(a)<0,
\]
and
\[
\lim_{\theta\downarrow 0}\left(
   \xi^{\hat{\alpha}}(\theta) \frac{F_a(\theta,\xi(\theta))}{
                           G_a(M,\theta,\xi(\theta))}
                           \exp\left[\xi^{\hat{\alpha}}(\theta)
                           \left(M
                   -J_\alpha(a)-K_a(M,\theta,\xi(\theta))\right)\right]
       \right)=0.
\]
Therefore, we deduce from \eqref{eq-liminf4} that
\begin{equation}
\label{eq-liminf5}
\liminf_{\theta\downarrow 0}
      \frac{\ln \E\left(e^{M\xi^{\hat{\alpha}-1}(\theta)\bar{Z}_\theta}
           \ind{\{\bar{Z}_\theta\geq \xi(\theta)\}}\right)}{
              \xi^{\hat{\alpha}}(\theta)}\geq
                          \lim_{\theta\downarrow 0}
                               K_a(M,\theta,\xi(\theta))=M^\alpha\eta_0I_\alpha.
\end{equation}
Now, it follows from H\"older's inequality that, for any $q>1$,
\begin{eqnarray*}
\E\left(e^{M\xi^{\hat{\alpha}-1}(\theta)\bar{Z}_\theta}
           \ind{\{\bar{Z}_\theta\geq \xi(\theta)\}}\right)
           &\leq &
           \left(\E\left(e^{qM\xi^{\hat{\alpha}-1}(\theta)\bar{Z}_\theta}
           \right)\right)^{1/q}
           \left[\P\left(\bar{Z}_\theta\geq \xi(\theta)\right)\right]^{1-\frac{1}{q}}\\
           &=&
           \exp\left[\frac{\theta}{q}\bar{\phi}\left(q
           M\xi^{\hat{\alpha}-1}(\theta)/\theta^{1/\alpha}\right)\right]
             \left[\P\left(\bar{Z}_\theta\geq \xi(\theta)\right)\right]^{1-\frac{1}{q}}.
\end{eqnarray*}
Hence
\begin{eqnarray*}
\left(1-\frac{1}{q}\right)\ln \P\left(\bar{Z}_\theta\geq \xi(\theta)\right)
&\geq&
\ln\left[\E\left(e^{M\xi^{\hat{\alpha}-1}(\theta)\bar{Z}_\theta}
           \ind{\{\bar{Z}_\theta\geq \xi(\theta)\}}\right)
\right]
 -\frac{\theta}{q}\bar{\phi}\left(q
           \frac{M\xi^{\hat{\alpha}-1}(\theta)}{\theta^{1/\alpha}}\right)\\
           &\geq&
         \ln\left[\E\left(e^{M\xi^{\hat{\alpha}-1}(\theta)\bar{Z}_\theta}
           \ind{\{\bar{Z}_\theta\geq \xi(\theta)\}}\right)
\right]\\
&&
 -\nu_a \theta^{1-\frac{1}{\alpha}}
 M\xi^{\hat{\alpha}-1}(\theta)
-M^\alpha\xi^{\hat{\alpha}}(\theta)q^{\alpha- 1}
    H_\alpha\left(
          \frac{qM\xi^{\hat{\alpha}-1}(\theta)}{\theta^{1/\alpha}}\right),
\end{eqnarray*}
where the last inequality follows from Lemma~\ref{lem-AS3}.
We now deduce from \eqref{eq-liminf5} and from the fact that
$\lim_{\rho\to \infty}H_\alpha(\rho)=\eta_0I_\alpha$
\[
\left(1-\frac{1}{q}\right)\liminf_{\theta\downarrow 0}
  \frac{\ln \P\left(\bar{Z}_\theta\geq \xi(\theta)\right)}{\xi^{\hat{\alpha}}(\theta)}
  \geq
      M^\alpha\eta_0I_\alpha \left(
         1-q^{\alpha- 1}\right).
\]
Hence
\[
\liminf_{\theta\downarrow 0}
\frac{\ln \P\left(\bar{Z}_\theta\geq \xi(\theta)\right)}{\xi^{\hat{\alpha}}(\theta)}
  \geq
      M^\alpha\eta_0I_\alpha\frac{q-q^\alpha}{
         q- 1},
\]
and, by taking the limit as $q$ goes to $1$,
\[
\liminf_{\theta\downarrow 0}
\frac{\ln \P\left(\bar{Z}_\theta\geq \xi(\theta)\right)}{\xi^{\hat{\alpha}}(\theta)}
  \geq
      -(\alpha-1)M^\alpha\eta_0I_\alpha\]
Since $M$ is arbitrary in
$\left((\alpha \eta_0 I_\alpha)^{\frac{-1}{\alpha-1}},+\infty\right)$,
we can take the limit as $M$ goes to
$(\alpha \eta_0 I_\alpha)^{\frac{-1}{\alpha-1}}$,
so that
\[
\liminf_{\theta\downarrow 0}
\frac{\ln \P\left(\bar{Z}_\theta\geq \xi(\theta)\right)}{\xi^{\hat{\alpha}}(\theta)}
  \geq
  -(\alpha-1)\frac{\eta_0I_\alpha}{
             (\alpha\eta_0 I_\alpha)^{\frac{\alpha}{\alpha-1}}}
             =-J_\alpha(0).
\]
\end{pf}

\noindent {\bf Proof of Proposition~\ref{prop-ASe}:}
We first prove
\begin{equation}
\label{eq-ASe1}
\liminf_{\theta\downarrow 0}
        \frac{\zeta(\theta)}{\theta^{1/\alpha}|\ln \theta|^{1-\frac{1}{\alpha}}}\geq
             \left(\alpha\eta_0 I_0\right)^{1/\alpha}.
\end{equation}
We deduce from \eqref{**'} that
\[
\E \left(e^{(r-\delta)\theta+X_\theta}
                                         -1-\zeta(\theta)\right)_+
                                         = (r-\delta)\theta +o(\theta).
\]
We have the decomposition $X_\theta=\bar{X}_\theta+X^0_\theta-\theta\int (e^y-1)_+\nu(dy)$,
where the processes $\bar{X}$ and $X^0$ are independent and
$\E\left(e^{X^0_\theta}\right)=e^{\theta\int (e^y-1)_+\nu(dy)}$, so that, by conditioning with respect
to $\bar{X}$,
\begin{eqnarray*}
\E \left(e^{(r-\delta)\theta+X_\theta}
                                         -1-\zeta(\theta)\right)_+
                                            &\geq&
                                            \E \left(e^{(r-\delta)\theta+\bar{X}_\theta}
                                         -1-\zeta(\theta)\right)_+\\
                                         &\geq&
                                        \E \left((r-\delta)\theta+\bar{X}_\theta
                                        -\zeta(\theta)\right)_+  \\
                                        &\geq&
                                        \E \left(\bar{X}_\theta
                                        -\zeta(\theta)\right)_+.
\end{eqnarray*}
Hence, with the notations of Lemma~\ref{lem-asymp},
\[
\E \left(\bar{Z}_\theta
                                        -\bar{\zeta}(\theta)\right)_+\leq (r-\delta)\theta^{1-\frac{1}{\alpha}}
                                                        +o(\theta^{1-\frac{1}{\alpha}}),
\]
where
\[
\bar{\zeta}(\theta)=\frac{\zeta(\theta)}{\theta^{1/\alpha}}.
\]
We deduce thereof that there exists a positive constant $C$ such that, for $\theta$ close to $0$,
\begin{equation}
\label{eq-ASe2}
\ln \E \left(\bar{Z}_\theta
                                        -\bar{\zeta}(\theta)\right)_+\leq \left(1-\frac{1}{\alpha}\right)\ln \theta +C.
\end{equation}
Now, given any $\beta>1$, we have
\[
\E \left(\bar{Z}_\theta
                                        -\bar{\zeta}(\theta)\right)_+\geq (\beta-1)\P\left(\bar{Z}_\theta
                                        \geq\beta \bar{\zeta}(\theta)\right),
\]
so that
\[
\ln \E \left(\bar{Z}_\theta
                                        -\bar{\zeta}(\theta)\right)_+\geq \ln (\beta-1)+
                                            \ln \P\left(\bar{Z}_\theta
                                        \geq\beta\bar{\zeta}(\theta)\right).
\]
Hence
\begin{eqnarray*}
\liminf_{\theta\downarrow 0}
    \frac{\ln \E \left(\bar{Z}_\theta
                                        -\bar{\zeta}(\theta)\right)_+}{\bar{\zeta}^{\frac{\alpha}{\alpha -1}}(\theta)}
                                        &\geq &
                                        \liminf_{\theta\downarrow 0}
                                         \frac{\ln \P\left(\bar{Z}_\theta
                                        \geq\beta\bar{\zeta}(\theta)\right)}{\bar{\zeta}^{\frac{\alpha}{\alpha -1}}(\theta)}\\
                                        &=&-\beta^{\frac{\alpha}{\alpha -1}}J_\alpha(0),
\end{eqnarray*}
where the last inequality follows from Lemma~\ref{lem-asymp}, applied with $\xi(\theta)=\beta\bar{\zeta}(\theta)$.
Going back to \eqref{eq-ASe2}, we deduce
\[
\left(1-\frac{1}{\alpha}\right)
\liminf_{\theta\downarrow 0}\frac{\ln(\theta)}{\bar{\zeta}^{\frac{\alpha}{\alpha -1}}(\theta)}
   \geq - \beta^{\frac{\alpha}{\alpha -1}}J_\alpha(0).
\]
Since $\beta$ is arbitrary in $(1,+\infty)$, we have
\[
\left(1-\frac{1}{\alpha}\right)
\limsup_{\theta\downarrow 0}\frac{|\ln(\theta)|}{\bar{\zeta}^{\frac{\alpha}{\alpha -1}}(\theta)}
   \leq J_\alpha(0).
\]
Therefore
\[
\liminf_{\theta\downarrow 0}\frac{\bar{\zeta}^{\frac{\alpha}{\alpha -1}}(\theta)}{|\ln(\theta)|}\geq
   \frac{\alpha-1}{\alpha J_\alpha(0)}=\left(\alpha\eta_0I_0\right)^{\frac{1}{\alpha-1}},
\]
which proves \eqref{eq-ASe1}.

In order to derive an upper bound for
$\limsup_{\theta\downarrow 0}
\frac{\zeta(\theta)}{\theta^{1/\alpha}|\ln \theta|^{1-\frac{1}{\alpha}}}$,
 we first deduce
from~\eqref{**'} a lower bound for
$\displaystyle\E\left(e^{\bar{X}_\theta}-1-\zeta(\theta)\right)_+$.
We have
\begin{eqnarray*}
\E\left(e^{(r-\delta)\theta+X_\theta}
                                         -1-\zeta(\theta)\right)_+&=&
                                         \E\left(
                                          e^{(r-\delta)\theta+X_\theta}
                                         \ind{\{X_\theta\geq \ln(1+\zeta(\theta))
                                         -(r-\delta)\theta\}}\right)\\
                                         &&-(1+\zeta(\theta))
                                         \P(X_\theta\geq \ln(1+\zeta(\theta))-
                                         (r-\delta)\theta).
\end{eqnarray*}
Note that
\[
\P(X_\theta\geq \ln(1+\zeta(\theta))-
                                         (r-\delta)\theta)=
          \P\left(\frac{X_\theta}{\theta^{1/\alpha}}\geq
          \frac{ \ln(1+\zeta(\theta))-
                                         (r-\delta)\theta}{\theta^{1/\alpha}}\right).
\]
Since $X_\theta/\theta^{1/\alpha}$ weakly
 converges to a finite random variable
$Z$ as $\theta\downarrow 0$ and
$\lim_{\theta\downarrow 0}\zeta(\theta)/\theta^{1/\alpha}=+\infty$, we have
\[
\lim_{\theta\downarrow 0}
  \P\left(\frac{X_\theta}{\theta^{1/\alpha}}\geq
          \frac{ \ln(1+\zeta(\theta))-
                                         (r-\delta)\theta}{\theta^{1/\alpha}}\right)=0.
\]
Note that we also have
\[
\lim_{\theta \downarrow 0}\E\left(e^{(r-\delta)\theta+X_\theta}
                                         -1-\zeta(\theta)\right)_+=0.
\]
Therefore
\[
\lim_{\theta \downarrow 0}\E \left(e^{X_\theta}
                                         \ind{\{X_\theta\geq \ln(1+\zeta(\theta))-
                                         (r-\delta)\theta\}}\right)=0
\]
and
\begin{eqnarray*}
\E\left(e^{(r-\delta)\theta+X_\theta}
                                         -1-\zeta(\theta)\right)_+&=&
                                         \E\left( e^{X_\theta}
                                         \ind{\{X_\theta\geq \ln(1+\zeta(\theta))-
                                         (r-\delta)\theta\}}\right)\\
                                         &&-(1+\zeta(\theta))
                                         \P(X_\theta\geq \ln(1+\zeta(\theta))-
                                         (r-\delta)\theta)+o(\theta),
\end{eqnarray*}
so that, using \eqref{**'},
\begin{eqnarray}
(r-\delta)\theta
                &=&
                         \E \left(e^{X_\theta}
                                         -(1+\zeta(\theta))\right)
                                         \ind{\{X_\theta\geq \ln(1+\zeta(\theta))-
                                         (r-\delta)\theta\}}+o(\theta)\nonumber\\
                                         &\leq&
                            \E \left(e^{X_\theta}
                                         -1-\zeta(\theta)\right)_+
                                        +o(\theta).\label{4}
\end{eqnarray}
Using the decomposition $X_\theta=\bar{X}_\theta+X^0_\theta
     -\theta\int(e^y-1)_+\nu(dy)$,
     the independence of $\bar{X}$ and $X^0$, and the equality
     $\E e^{\bar{X}_\theta}=1$ (which follows from Lemma~\ref{lem-AS3}),
     we have
 \begin{eqnarray*}
\E \left(e^{X_\theta}-1-\zeta(\theta)\right)_+
     &\leq &
            \E\left(e^{X^0_\theta+\bar{X}_\theta}-1-\zeta (\theta)\right)_+\\
     &=&
          \E\left(\left(e^{X^0_\theta}-1\right)e^{\bar{X}_\theta}
              +e^{\bar{X}_\theta}-1-\zeta (\theta)\right)_+\\
     &\leq &
     \E\left(\left(e^{X^0_\theta}-1\right)e^{\bar{X}_\theta}\right)+
     \E\left(e^{\bar{X}_\theta}-1-\zeta (\theta)\right)_+\\
     &=&e^{\theta\int (e^y-1)_+\nu(dy)}-1 +
          \E\left(e^{\bar{X}_\theta}-1-\zeta (\theta)\right)_+\\
          &=&\theta\int (e^y-1)_+\nu(dy)+
             \E\left(e^{\bar{X}_\theta}-1-\zeta (\theta)\right)_+ +o(\theta).
\end{eqnarray*}
Hence, going back to (\ref{4}),
\ben\label{5}
\left(r-\delta-\int (e^y-1)_+\nu(dy)\right)\theta
        \leq   \E\left(e^{\bar{X}_\theta}-1-\zeta (\theta)\right)_+ +o(\theta).
\een
Introducing the notation $l(\theta)=\ln(1+\zeta(\theta))$, we have
\begin{eqnarray*}
 \E\left(e^{\bar{X}_\theta}-1-\zeta (\theta)\right)_+ &=&
       \E\left(e^{\bar{X}_\theta}-e^{l(\theta)}\right)_+ \\
       &=&
        \E\left(\ind{\{\bar{X}_\theta\geq l(\theta)\}}
              \int_{l(\theta)}^{\bar{X}_\theta}e^y dy\right)_+\\
              &\leq&
              \int_{l(\theta)}^{+\infty}
                e^y\P\left(\bar{X}_\theta\geq y\right)dy\\
                &=&
                \theta^{1/\alpha}
                 \int_{\bar{l}(\theta)}^{+\infty}
                e^{z\theta^{1/\alpha}}\P\left(\bar{X}_\theta\geq
                                     z\theta^{1/\alpha}\right)dz,
\end{eqnarray*}
with $\bar{l}(\theta)=l(\theta)/\theta^{1/\alpha}$.
It follows fromLemma~\ref{lem-tail} that, given any $a\in [a_0,0)$,
we have
\[
\P\left(\bar{X}_\theta\geq
                                     z\theta^{1/\alpha}\right)
  \leq \exp\left(C_a\theta^{1-\frac{1}{\alpha}} z^{\frac{1}{\alpha-1}}
       -J_\alpha(a)z^{\frac{\alpha}{\alpha-1}}\right).
\]
Now, fix $\epsilon>0$. Since
$\lim_{\theta\downarrow 0}\bar{l}(\theta)=+\infty$, we have, for
$\theta$ close to $0$,
\[
\forall z\geq \bar{l}(\theta),\quad
z\theta^{1/\alpha}+C_a\theta^{1-\frac{1}{\alpha}} z^{\frac{1}{\alpha-1}}
   \leq \epsilon z^{\frac{\alpha}{\alpha-1}}.
\]
Hence, with the notation $\hat{\alpha}=\frac{\alpha}{\alpha-1}$,
\begin{eqnarray*}
 \E\left(e^{\bar{X}_\theta}-1-\zeta (\theta)\right)_+ &\leq&
   \theta^{1/\alpha}
                 \int_{\bar{l}(\theta)}^{+\infty}
                e^{-(J_\alpha(a)-\epsilon)z^{\hat{\alpha}}}dz.
\end{eqnarray*}
We can assume $\epsilon$ close enough to $0$ so that
 $J_\alpha(a)>\epsilon$, and
\begin{eqnarray*}
 \int_{\bar{l}(\theta)}^{+\infty}
                e^{-(J_\alpha(a)-\epsilon)z^{\hat{\alpha}}}dz
               & \leq&
                \frac{1}{\hat{\alpha}(J_\alpha(a)-\epsilon)
                      (\bar{l}(\theta))^{\hat{\alpha}-1}}
                 \int_{\bar{l}(\theta)}^{+\infty}
                 \hat{\alpha}(J_\alpha(a)-\epsilon)
                 z^{\hat{\alpha}-1}
                e^{-(J_\alpha(a)-\epsilon)z^{\hat{\alpha}}}dz\\
                &=&
                \frac{1}{\hat{\alpha}(J_\alpha(a)-\epsilon)
                      (\bar{l}(\theta))^{\hat{\alpha}-1}}
                      e^{-(J_\alpha(a)-\epsilon)\bar{l}^{\hat{\alpha}}(\theta)}.
\end{eqnarray*}
Going back to \eqref{5}, we deduce
\[
\left(r-\delta-\int (e^y-1)_+\nu(dy)\right)\theta^{1-\frac{1}{\alpha}}
    \leq  \frac{1}{\hat{\alpha}(J_\alpha(a)-\epsilon)
                      (\bar{l}(\theta))^{\hat{\alpha}-1}}
                      e^{-(J_\alpha(a)-\epsilon)\bar{l}^{\hat{\alpha}}(\theta)}
                      +o(\theta^{1-\frac{1}{\alpha}}),
\]
so that, for $\theta$ close to $0$,
\begin{eqnarray*}
\theta^{1-\frac{1}{\alpha}}\leq \frac{C_\epsilon}{(\bar{l}(\theta))^{\hat{\alpha}-1}}
         e^{-(J_\alpha(a)-\epsilon)\bar{l}^{\hat{\alpha}}(\theta)},
\end{eqnarray*}
where $C_\epsilon $ is a positive constant.
Hence
\[
\left(1-\frac{1}{\alpha}\right)\ln \theta \leq -(J_\alpha(a)-\epsilon)\bar{l}^{\hat{\alpha}}(\theta)+
        \ln \left(\frac{C_\epsilon}{(\bar{l}(\theta))^{\hat{\alpha}-1}}\right),
\]
and
\[
\left(1-\frac{1}{\alpha}\right)\limsup_{\theta\downarrow 0}
            \frac{\ln \theta}{\bar{l}^{\hat{\alpha}}(\theta)}
               \leq -(J_\alpha(a)-\epsilon).
\]
Since $a$ and $\epsilon$ can be arbitrarily close to $0$, we get, in the limit,
\[
\limsup_{\theta\downarrow 0}
            \frac{\ln \theta}{\bar{l}^{\hat{\alpha}}(\theta)}
   \leq - \frac{\alpha}{\alpha-1}J_\alpha(0)=-(\alpha\eta_0I_0)^{\frac{-1}{\alpha-1}}.
\]
Note that $\lim_{\theta\downarrow 0}\frac{l(\theta)}{\zeta(\theta)}=1$, so that we can conclude that
\[
\limsup_{\theta\downarrow 0}
            \frac{\bar{\zeta}^{\hat{\alpha}}(\theta)}{|\ln \theta|}
   \leq (\alpha\eta_0I_0)^{\frac{1}{\alpha-1}}.
\]
\qed

 \subsection{Estimating the difference $b_e-b$}
 \begin{prop}\label{Prop-StableAmEu}
   Under assumption (AS), we have
   \[
\limsup_{t\to T}\frac{b_e(t)-b(t)}{(T-t)^{1/\alpha}}<\infty.
\]
 \end{prop}
 \begin{pf}
 It follows from the variational inequality and the inequality
 $\partial P/\partial t \leq 0$
  that, for $t\in (0,T)$
 and $x\in(b(t),K)$, we have
 \[
 (r-\delta)x\frac{\partial P}{\partial x}(t,x)+
      \int\left(P(t,xe^y)-P(t,x)-x\frac{\partial P}{\partial x}(t,x)(e^y-1)
              \right)\nu(dy)\geq r(K-x).
 \]
 Note that, since $\int (e^y-1)_+\nu(dy)<\infty$, we may write
 \begin{eqnarray*}
\int_{(0,+\infty)}\left(P(t,xe^y)-P(t,x)-x\frac{\partial P}{\partial x}(t,x)(e^y-1)
              \right)\nu(dy)&=&
              \int_{(0,+\infty)}\left(P(t,xe^y)-P(t,x)\right)\nu(dy)\\
             &&- x\frac{\partial P}{\partial x}(t,x)\int_{(0,+\infty)}\left(e^y-1
                     \right)\nu(dy),
\end{eqnarray*}
so that, using the notation
\[
\gamma_+=r-\delta-\int\left(e^y-1
                     \right)_+\nu(dy),
\]
we get
\begin{eqnarray*}
\lefteqn{
\gamma_+x\frac{\partial P}{\partial x}(t,x)
+ \int_{(0,+\infty)}\left(P(t,xe^y)-P(t,x)\right)\nu(dy)}\\&&
   +\int_{(-\infty,0)}\left(P(t,xe^y)-P(t,x)-x\frac{\partial P}{\partial x}(t,x)(e^y-1)
              \right)\nu(dy)\geq r(K-x).
\end{eqnarray*}
Therefore, for $x\in(b(t),K)$,
\begin{eqnarray*}
\int_{(-\infty,0)}\left(P(t,xe^y)-P(t,x)-x\frac{\partial P}{\partial x}(t,x)(e^y-1)
              \right)\nu(dy)&\geq
              &
               -\gamma_+x\frac{\partial P}{\partial x}(t,x)
              -J(t),
\end{eqnarray*}
where
\[
J(t)=\int_{(0,+ \infty)}\;
          \sup_{b(t)<x<K}\left|P(t,xe^y)-P(t,x)\right|\,\nu(dy).
\]
Note that, due to the  Lipschitz property of $P(t,.)$, we have,
for $x\in (b(t),K)$ and $y>0$,
\begin{eqnarray*}
0\leq P(t,xe^y)-P(t,x)
         &\leq &x\left(e^y-1\right)\\
         &\leq &K\left(e^y-1\right),
\end{eqnarray*}
and
\[
P(t,xe^y)-P(t,x)\leq P(t,b(t)).
\]
Since $\lim_{t\to T}P(t,b(t))=P(T,K)=0$, we deduce
\ben\label{eq-diff1}
\lim_{t\to T}\int_{(0,+\infty)}\,
          \sup_{b(t)<x<K}\left|P(t,xe^y)-P(t,x)\right|\nu(dy)=0.
\een
Now, for $x\in(b(t),b_e(t))$, we have (with $\theta=T-t$)
\[
\frac{\partial P}{\partial x}(t,x)
       \leq \frac{\partial_- P}{\partial x}(t,b_e(t))\leq
             \frac{\partial_- P_e}{\partial x}(t,b_e(t))=
             -\E \left(e^{-\delta\theta+X_\theta}
             \ind{\left\{b_e(t)e^{(r-\delta)\theta+X_\theta}\leq K\right\}}\right),
\]
where $\partial_-$ refers to left-hand derivatives,
the first inequality follows from the convexity of
$P(,t,.)$ and the second inequality follows from the fact that
$x\mapsto (P-P_e)(t,x)$ is non-increasing (see Corollary~\ref{cor-eep}).
Observe that
\begin{eqnarray*}
\E \left(e^{-\delta\theta+X_\theta}
             \ind{\left\{b_e(t)e^{(r-\delta)\theta+X_\theta}\leq K\right\}}\right)
             &=&\E \left(e^{-\delta\theta+X_\theta}
             \ind{\left\{(r-\delta)\theta+X_\theta\leq \ln(1+\zeta(\theta))\right\}}
                            \right)\\
             &=&
             \E \left(e^{-\delta\theta+X_\theta}
             \ind{\left\{\frac{(r-\delta)\theta+X_\theta}{\theta^{1/\alpha}}
                   \leq \frac{\ln(1+\zeta(\theta))}{\theta^{1/\alpha}}\right\}}
                            \right)
\end{eqnarray*}
Using (\ref{eq-AS2.1}) and Lemma~\ref{lem-AS1}, we derive
\[
\lim_{\theta\to 0}
   \E \left(e^{-\delta\theta+X_\theta}
             \ind{\left\{\frac{(r-\delta)\theta+X_\theta}{\theta^{1/\alpha}}
                   \leq \frac{\ln(1+\zeta(\theta))}{\theta^{1/\alpha}}\right\}}
                            \right)=1.
\]
Now, for $x\in (b(t),K)$ denote
\[
I(t,x)=\int_{(-\infty,0)}\left(P(t,xe^y)-P(t,x)-x\frac{\partial P}{\partial x}(t,x)(e^y-1)
              \right)\nu(dy).
\]
It follows from the above discussion that
\ben\label{eq-lowI}
\liminf_{t\to T}\inf_{x\in (b(t),b_e(t))}I(t,x)
      \geq \gamma_+ K.
\een
We will now derive an upper bound for $I(t,x)$, for $b(t)<x<K$. We have
\begin{eqnarray*}
I(t,x)&=&\int_{\left(-\infty,\ln {b(t)\over x}\right]}
     \left(P(t,xe^y)-P(t,x)-x\frac{\partial P}{\partial x}(t,x)(e^y-1)
              \right)\nu(dy)\\
              &&
             \;\;\; +\int_{\left(\ln {b(t)\over x}, 0\right)}
     \left(P(t,xe^y)-P(t,x)-x\frac{\partial P}{\partial x}(t,x)(e^y-1)
              \right)\nu(dy).
\end{eqnarray*}
For $y\leq\ln {b(t)\over x}$, we have
\begin{eqnarray*}
  P(t,xe^y)-P(t,x)-x\frac{\partial P}{\partial x}(t,x)(e^y-1)
  &=&(K-xe^y)-P(t,x)-x\frac{\partial P}{\partial x}(t,x)(e^y-1)\\
  &\leq&(K-xe^y)-(K-x)-x\frac{\partial P}{\partial x}(t,x)(e^y-1)\\
  &=&x\left(1+\frac{\partial P}{\partial x}(t,x)\right)\left(1-e^y\right).
\end{eqnarray*}
For $y\in (\ln(b(t)/x),0)$, we have, using the convexity of $P(t,.)$,
\begin{eqnarray*}
  P(t,xe^y)-P(t,x)-x\frac{\partial P}{\partial x}(t,x)(e^y-1)
  &\leq&x(e^y-1)\frac{\partial P}{\partial x}(t,xe^y)
          -x\frac{\partial P}{\partial x}(t,x)(e^y-1)\\
  &=&x\left(\frac{\partial P}{\partial x}(t,x)-
       \frac{\partial P}{\partial x}(t,xe^y)\right)\left(1-e^y\right).
\end{eqnarray*}
Therefore
\begin{eqnarray}
I(t,x)&\leq&
   x\left(1+\frac{\partial P}{\partial x}(t,x)\right)
            \int_{\left(-\infty,\ln {b(t)\over x}\right]}\left(1-e^y\right)\nu(dy)\nonumber\\
          &&  +
            \int_{\left(\ln {b(t)\over x}, 0\right)}
   x\left(\frac{\partial P}{\partial x}(t,x)-
       \frac{\partial P}{\partial x}(t,xe^y)\right)\left(1-e^y\right)\nu(dy).\label{eq-highI1}
\end{eqnarray}
Due to (\ref{eq-lowI}), there exists $\eta >0$ such that for $t\in [T-\eta,T)$,
\ben\label{eq-highI2}
\inf_{x\in (b(t),b_e(t))}I(t,x)
      \geq \frac{\gamma_+ K}{2}.
\een
From now on, we assume $t\in [T-\eta, T)$ and, for
$\xi\in\left(0,\ln(b_e(t)/b(t))\right)$ we set
\[
g_t(\xi)=P(t,b(t)e^\xi).
\]
Note that the derivative of $g_t$ is given by
\[
g'_t(\xi)=b(t)e^\xi{\partial P\over \partial x}(t,b(t)e^\xi),
\]
and, due to the smooth fit property, $g'_t(0)=-b(t)$.
We also have $|g'_t(\xi)|\leq b_e(t)\leq K$.
Applying (\ref{eq-highI1}) with $x=b(t)e^\xi$, we have,
using~(\ref{eq-highI2}),
\begin{eqnarray*}
 \frac{\gamma_+ K}{2}&\leq &
 \left(g'_t(\xi)-g'_t(0)e^\xi\right)\int_{(-\infty,-\xi]}\left(1-e^y\right)\nu(dy)\\
 &&
  \int_{(-\xi, 0)}\left(g'_t(\xi)-g'_t(\xi+y)e^{-y}\right)\left(1-e^y\right)\nu(dy)
\end{eqnarray*}
Note that
\begin{eqnarray*}
\left(g'_t(\xi)-g'_t(0)e^\xi\right)\int_{(-\infty,-\xi]}\left(1-e^y\right)\nu(dy)
&=&\left(g'_t(\xi)-g'_t(0)\right)\int_{(-\infty,-\xi]}\left(1-e^y\right)\nu(dy)\\
&&
             +g'_t(0)(1-e^\xi)\int_{(-\infty,-\xi]}\left(1-e^y\right)\nu(dy).
\end{eqnarray*}
For $\xi \in \left(0,\ln(b_e(t)/b(t))\right)$, we have, for any $\epsilon>0$,
\begin{eqnarray*}
\left(e^\xi-1\right)\int_{(-\infty,-\xi]}\left(1-e^y\right)\nu(dy)&\leq&
    \left(e^\xi-1\right)\nu\left((-\infty,-\epsilon]\right)+
                  \int_{(-\epsilon,0)}\left(e^{-y}-1\right)\left(1-e^y\right)\nu(dy) \\
                  &\leq&
                  \left({b_e(t)\over b(t)}-1\right)\nu\left((-\infty,-\epsilon]\right)+
                  \int_{(-\epsilon,0)}\left(e^{-y}-1\right)\left(1-e^y\right)\nu(dy).
\end{eqnarray*}
Therefore
\[
\lim_{t\to T}\sup_{\xi\in\left(0,\ln(b_e(t)/b(t))\right)}
      \left(e^\xi-1\right)\int_{(-\infty,-\xi]}\left(1-e^y\right)\nu(dy)=0.
\]
By taking $\eta$ smaller if necessary, we can now assume that,
for $t\in [T-\eta, T)$ and $\xi \in  \left(0,\ln(b_e(t)/b(t))\right)$, we have
\begin{eqnarray}
\frac{\gamma_+ K}{3}&\leq &
 \left(g'_t(\xi)-g'_t(0)\right)\int_{(-\infty,-\xi]}\left(1-e^y\right)\nu(dy)
                   \nonumber\\
 &&
  \int_{(-\xi, 0)}\left(g'_t(\xi)-g'_t(\xi+y)e^{-y}\right)\left(1-e^y\right)\nu(dy)
      \label{eq-highI**}.
\end{eqnarray}
Now, take $a\in  \left(0,\ln(b_e(t)/b(t))\right)$. By integrating
(\ref{eq-highI**}) with respect to $\xi$ from $0$ to $a$, we get
\[
\frac{\gamma_+ K}{3}a
   \leq j_1(a)+j_2(a),
\]
where
\[
j_1(a)=\int_0^a d\xi
      \left(g'_t(\xi)-g'_t(0)\right)\left(\int_{(-\infty,-\xi]}\left(1-e^y\right)\nu(dy)\right)
\]
 and
\[
j_2(a)=\int_0^a d\xi
     \int_{(-\xi, 0)}\nu(dy)\left(g'_t(\xi)-g'_t(\xi+y)e^{-y}\right)\left(1-e^y\right).
\]
In order to estimate $j_1(a)$ we note that, for $\xi\in(0,a)$,
\begin{eqnarray*}
  g'_t(\xi)&=&b(t)e^\xi\frac{\partial P}{\partial x}(t,b(t)e^\xi)\\
              &\leq &b(t)e^\xi\frac{\partial P}{\partial x}(t,b(t)e^a)\\
              &=&e^{\xi -a}g'_t(a)\leq e^{-a}g'_t(a),
\end{eqnarray*}
where the first inequality follows from the convexity of $P(t,.)$ and the
second one from $g'_t(a)\leq 0$.
Hence
\begin{eqnarray*}
j_1(a)&\leq &\left(e^{-a}g'_t(a)-g'_t(0)\right)
                 \int_0^a d\xi
      \left(\int_{(-\infty,-\xi]}\left(1-e^y\right)\nu(dy)\right) \\
      &=&
           \left(e^{-a}g'_t(a)-g'_t(0)\right)
           \int_{-\infty}^0\nu(dy)\left(1-e^y\right)
                   \left(\int_0^{a\wedge (-y)} d\xi\right).
\end{eqnarray*}
Note that
$e^{-a}g'_t(a)-g'_t(0)=b(t)\left(1+\frac{\partial P}{\partial x}(t,b(t)e^a)
        \right)\geq 0$.
Using the assumptions we have on $\nu$, we can find $\beta<0$
such that, for $y\in (\beta, 0)$,
\[
\nu(dy)\leq \frac{2\kappa}{|y|^{1+\alpha}}dy,
\]
so that, using $e^y\geq 1+y$,
\begin{eqnarray*}
j_1(a)&\leq &
    \left(e^{-a}g'_t(a)-g'_t(0)\right)
    \left(a \int_{(-\infty,\beta]}\nu(dy)\left(1-e^y\right)
                   +
                    \int_{(\beta,0)}\frac{2\kappa dy}{|y|^{1+\alpha}}\left(1-e^y\right)
                   \left(\int_0^{(a\wedge-y)} d\xi\right)
                   \right)\\
                   &\leq&
         \left(e^{-a}g'_t(a)-g'_t(0)\right)
    \left(a \nu((-\infty,\beta])
                   +
                    \int_0^a d\xi
                    \int_\xi^{|\beta|}\frac{2\kappa dy}{|y|^{1+\alpha}}|y|
                                      \right)     \\
                                      &\leq &
            \left(e^{-a}g'_t(a)-g'_t(0)\right)
    \left(a \nu((-\infty,\beta])
                   +
                   \frac{ 2\kappa}{\alpha-1}\int_0^a \xi^{1-\alpha}d\xi
                                      \right)      \\
                                      &=&
                                                                   \left(e^{-a}g'_t(a)-g'_t(0)\right)
    \left(a \nu((-\infty,\beta])
                   +
                   \frac{ 2\kappa}{(2-\alpha)(\alpha-1)}a^{2-\alpha}
                                      \right)
\end{eqnarray*}
Note that $a \in  \left(0,\ln(b_e(t)/b(t))\right)$ and
 $\lim_{t\to T}\ln(b_e(t)/b(t))=0$. So, for $t$ close
 enough to $T$, we may assume $a\in(0,1]$, so that $a\leq a^{2-\alpha}$
 (recall $1<\alpha<2$). Therefore, for some $C>0$,
 \begin{eqnarray}
 j_1(a)&\leq &C a^{2-\alpha}\left(e^{-a}g'_t(a)-g'_t(0)\right)\nonumber\\
          &=& Ca^{2-\alpha}\left(g'_t(a)-g'_t(0)\right)+
          Cg'_t(a)a^{2-\alpha}\left(e^{-a}-1\right)\nonumber\\
          &\leq &Ca^{2-\alpha}\left(g'_t(a)-g'_t(0)\right)+
          CKa^{3-\alpha}.  \label{eq-j1}
 \end{eqnarray}
 We now study $j_2(a)$. Note that, for $y<0$,
\begin{eqnarray*}
 g'_t(\xi)-g'_t(\xi+y)e^{-y}
    &= &b(t)e^{\xi}\left(\frac{\partial P}{\partial x}\left(t, b(t)e^\xi\right)
        -\frac{\partial P}{\partial x}\left(t,b(t)e^{\xi+y}\right)\right)\geq 0.
 \end{eqnarray*}
 Since $a \in  \left(0,\ln(b_e(t)/b(t))\right)$ and
 $\lim_{t\to T}\ln(b_e(t)/b(t))=0$, we may assume $a<|\beta|$ and
 write
 \begin{eqnarray*}
j_2(a)&\leq &\int_0^a d\xi
            \int_{(-\xi, 0)}\frac{2\kappa}{|y|^{1+\alpha}}dy\left(g'_t(\xi)-g'_t(\xi+y)e^{-y}\right)\left(1-e^y\right)\\
              &=&
              2\kappa
               \int_0^a d\xi
            \int_0^\xi \frac{dy}{y^{1+\alpha}}\left(g'_t(\xi)-g'_t(\xi-y)e^{y}\right)\left(1-e^{-y}\right)\\
            &\leq &
            2\kappa
               \int_0^a d\xi
            \int_0^\xi \frac{dy}{y^{\alpha}}\left(g'_t(\xi)-g'_t(\xi-y)e^{y}\right),
            \end{eqnarray*}
where  the last inequality follows from $1-e^{-y}\leq y$.
Hence
\begin{eqnarray*}
j_2(a)&\leq &
   2\kappa
               \int_0^a\frac{dy}{y^{\alpha}}
            \int_y^a d\xi\left(g'_t(\xi)-g'_t(\xi-y)e^{y}\right)\\
            &=&
            2\kappa
               \int_0^a\frac{dy}{y^{\alpha}}
            \left(g_t(a)-g_t(a-y)e^y-g_t(y)+g_t(0)e^y\right).\\
            &=&
            2\kappa
               \int_0^a\frac{dy}{y^{\alpha}}
            \left(g_t(a)-g_t(a-y)-g_t(y)+g_t(0)\right)
              + 2\kappa
               \int_0^a\frac{dy}{y^{\alpha}}( e^y-1)\left(g_t(0)-g_t(a-y)\right)
              \\
              &\leq &
             2\kappa
               \int_0^a\frac{dy}{y^{\alpha}}
            \left(g_t(a)-g_t(a-y)-g_t(y)+g_t(0)\right)
            +  2\kappa Ka \int_0^a\frac{dy}{y^{\alpha}}( e^y-1),
\end{eqnarray*}
where the last inequality follows from $||g'_t||_\infty \leq K$.
Note that $a \int_0^a\frac{dy}{y^{\alpha}}( e^y-1)\leq C a^{3-\alpha}$
for some $C>0$. On the other hand,
we have, for $y\in (0,a)$,
\begin{eqnarray*}
g_t(a)-g_t(a-y)&=&\int_0^yg'_t(a-z)dz\\
      &=&\int_0^y b(t)e^{a-z}
           \frac{\partial P}{\partial x}\left(t,b(t)e^{a-z}\right)dz\\
           &\leq&
              \int_0^y b(t)e^{a-z}
           \frac{\partial P}{\partial x}\left(t,b(t)e^{a}\right)dz\\
           &\leq &
           \int_0^y b(t)
           \frac{\partial P}{\partial x}\left(t,b(t)e^{a}\right)dz\\
           &=&yb(t)\frac{\partial P}{\partial x}\left(t,b(t)e^{a}\right)
             =ye^{-a}g'_t(a),
\end{eqnarray*}
where the first inequality follows from the convexity of $P(t,.)$ and the
second one from $\partial P/\partial x\leq 0$.
Similarly, we have
\begin{eqnarray*}
g_t(y)-g_t(0)&=&\int_0^yg'_t(z)dz\\
      &=&\int_0^y
      b(t)e^{z}
           \frac{\partial P}{\partial x}\left(t,b(t)e^{z}\right)dz\\
           &\geq&
           \int_0^y
      b(t)e^{z}
           \frac{\partial P}{\partial x}\left(t,b(t)\right)dz\\
           &\geq&yb(t)e^y\frac{\partial P}{\partial x}\left(t,b(t)\right)=
             g'_t(0)ye^y.
\end{eqnarray*}
Hence
\begin{eqnarray*}
\int_0^a\frac{dy}{y^{\alpha}}
            \left(g_t(a)-g_t(a-y)-g_t(y)+g_t(0)\right)
            &\leq &
           \int_0^a\frac{dy}{y^{\alpha}} y\left(e^{-a}g'_t(a)-e^{y}g'_t(0)\right)\\
           &=&
     \int_0^a\frac{dy}{y^{\alpha-1}} \left(g'_t(a)-g'_t(0)\right) \\
     &&
     +     \int_0^a\frac{dy}{y^{\alpha-1}}\left[\left(e^{-a}-1\right)g'_t(a)
     +\left(1-e^y\right)g'_t(0)\right]\\
     &\leq &
     \int_0^a\frac{dy}{y^{\alpha-1}} \left(g'_t(a)-g'_t(0)\right) \\
     &&  +(a+(e^a-1))K\frac{a^{2-\alpha}}{2-\alpha}\\
     &\leq & C a^{2-\alpha} \left(g'_t(a)-g'_t(0)\right)
       +C a^{3-\alpha},
\end{eqnarray*}
for some $C>0$, so that we have
\ben\label{eq-j2}
j_2(a)\leq C a^{2-\alpha} \left(g'_t(a)-g'_t(0)\right)
       +C a^{3-\alpha}
\een
Putting (\ref{eq-j1}) and (\ref{eq-j2}) together, we conclude that,
for some positive constant $C$, we have
\[
\frac{\gamma_+ K}{3}a
   \leq C a^{2-\alpha} \left(g'_t(a)-g'_t(0)\right)
       +C a^{3-\alpha}
\]
or, equivalently,
\[
\frac{\gamma_+ K}{3C}a^{\alpha-1}
         \left(1-\frac{3C}{\gamma_+ K}a^{2-\alpha}\right)
   \leq  g'_t(a)-g'_t(0).
\]
For $t$ close enough to $T$, we have, for all
$a\in  \left(0,\ln(b_e(t)/b(t))\right)$,
$a^{2-\alpha}<\frac{\gamma_+ K}{6C}$, hence
\[
\frac{\gamma_+ K}{6C}a^{\alpha-1}
   \leq  g'_t(a)-g'_t(0).
\]
We now integrate this inequality with respect to $a$ from $0$
to $a_t=\ln(b_e(t)/b(t))$ to derive
\[
a_t^\alpha \leq C\left(g_t(a_t)-a_tg'_t(0)-g_t(0)\right),
\]
where $C$ is a positive constant.
Hence
\begin{eqnarray*}
\frac{1}{C}(b_e(t)-b(t))^{\alpha}&\leq&
           P(t,b_e(t))+b(t)\ln {b_e(t)\over b(t)}-P(t,b(t))\\
           &\leq &
           P(t,b_e(t))+b_e(t)-b(t)-(K-b(t))\\
           &=&P(t,b_e(t))-P_e(t,b_e(t))\\
           &\leq &rK(T-t),
\end{eqnarray*}
where the last inequality follows from the Early Exercise Premium
Formula.
  \end{pf}

\end{document}